%% file: IsoChapter.tex
\documentclass[preprint2]{proto}
\usepackage{times}
\usepackage{xcolor}
\usepackage{graphicx}
\usepackage{wasysym}
\usepackage{float}
\usepackage{dblfloatfix}
\usepackage[switch]{lineno}

\graphicspath{{./}{Figures/}}

\def\mearth{{\rm\,M_\oplus}}

\newcommand{\Ou}{1I/`Oumuamua}

\begin{document}

\title{\textbf{\LARGE INTERSTELLAR OBJECTS AND EXOCOMETS}}

\author {\textbf{\large Alan Fitzsimmons}}
\affil{\small\em Astrophysics Research Centre, Queen's University Belfast, Belfast BT7 1NN, UK}

\author {\textbf{\large Karen Meech}}
\affil{\small\em Institute for Astronomy, 2680 Woodlawn Drive, Honolulu HI 96822}

\author {\textbf{\large Luca Matr\`a}}
\affil{\small\em School of Physics, Trinity College Dublin, The University of Dublin, College Green, Dublin 2, Ireland}

\author {\textbf{\large Susanne Pfalzner}}
\affil{\small\em Forschungszentrum J\"ulich, Wilhelm-Johnen-Stra\ss e, 52425 J\"ulich, Germany}

\begin{abstract}

\begin{list}{ } {\rightmargin 1in}
\baselineskip = 11pt
\parindent=1pc
{\small 
In this chapter we review our knowledge of our galaxy's cometary population outside our Oort Cloud - exocomets and Interstellar Objects (ISOs).   We start with a brief overview of planetary system formation, viewed as a general process around stars. We then take a more detailed look at the creation and structure of exocometary belts, as revealed by the unprecedented combination of theoretical and observational advances in recent years. The existence and characteristics of individual exocomets orbiting other stars is summarized, before looking at the mechanisms by  which they may be ejected into interstellar space. The discovery of the first two ISOs is then described, along with the surprising differences in their observed characteristics. We end by looking ahead to what advances may take place in the next decade. 
\\~\\~\\~}
\end{list}
\end{abstract}  

\section{\textbf{INTRODUCTION}}
\label{sec:intro}

\begin{figure}[ht!]
\begin{center}
\includegraphics[width=8.0cm]{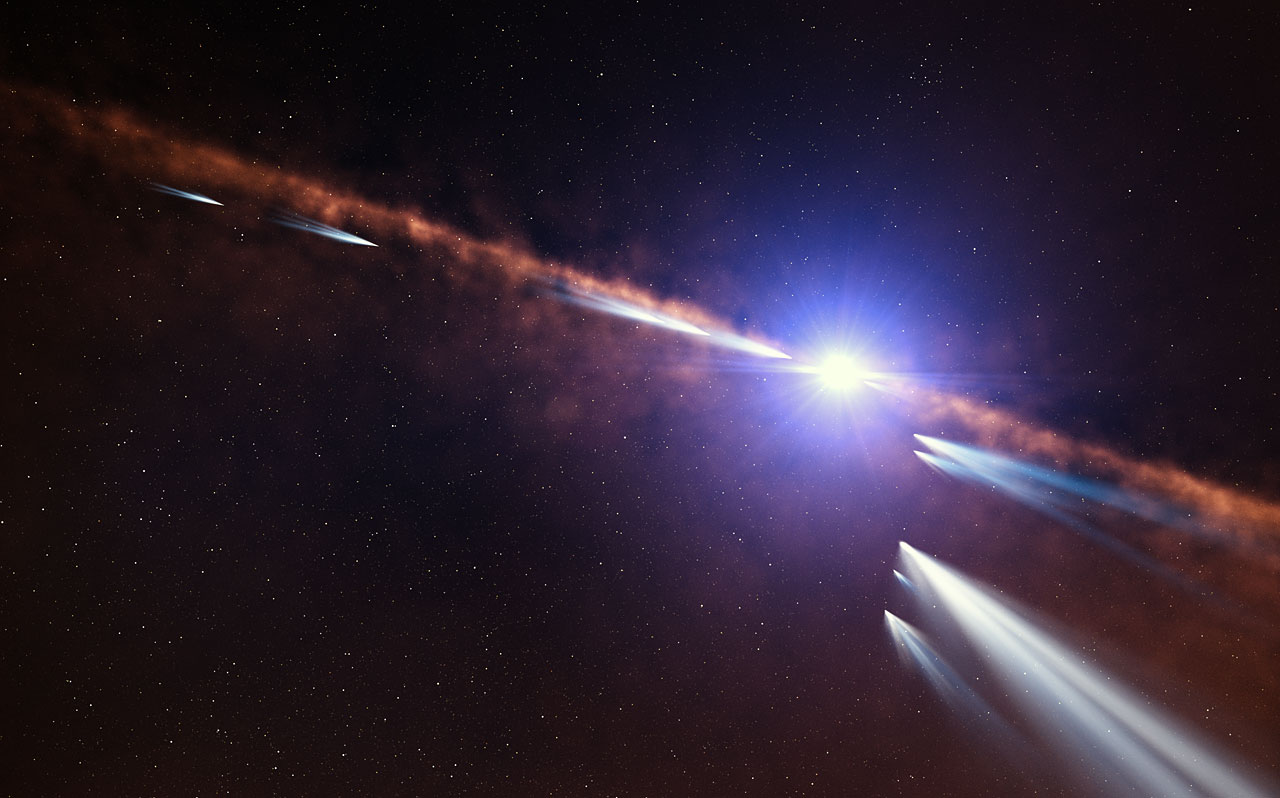}
\caption{Artist's impression of the $\beta$ Pic system, illustrating the extensive families of exocomets observed within this system (credit ESO/L. Cal\c{c}ada). }
\label{fig:BPicISO}
\vspace{-0.25cm}
\end{center}
\end{figure}

At the time of publication of {\it Comets II} \citep{Festou:2004}, the presence of cometary bodies around other stars had been well established. Since that time there has been a significant increase in theoretical investigations of cometary formation in protoplanetary disks (see \S\ref{sec:planetform:dust_growth} and {\it Simon et al., this volume}).
Resolved imaging and spectroscopy of circumstellar disks has become almost commonplace, giving exquisite insights into the 
birthplaces of comets  (see \S\ref{sec:planetform}).
Increasingly common detections of gas in extrasolar Kuiper belts has opened a new approach to study exocometary gas and composition, combined with a new era of dust detection at IR and sub-mm wavelengths. 
At the same time, observational studies have resulted in the discovery of new exocomet host systems, the detection of dust continuum transits for the first time, and further understanding of extant systems like $\beta$ Pictoris (hereafter $\beta$ Pic; see Fig.~\ref{fig:BPicISO} and \S\ref{sec:innerexocomets}).
White dwarf atmospheric modelling has revealed a 
variety of compositions of extrasolar small bodies, including potentially volatile-rich exocomets.

\textcolor{black}{Interstellar Objects (ISOs) are planetesimals unbound to stars, and generally thought to originate by ejection from their planetary systems by various physical mechanisms. These could be both asteroids that are generally considered to be inert rocky bodies, or  comets that contain a large amount of ices which can lead to outgassing and mass loss, the most common method of telling the difference between them (however, see the chapter by Jewitt \& Hsieh in this volume on the blurred boundaries between these objects). Assuming the same physics and dynamics occurs in other planetary systems as in our own, then the observable ISO population would primarily consist of cometary bodies of similar sizes to those orbiting our Sun, although see the discussion in \S\ref{sec:smallISOs}.}

\textcolor{black}{In contrast to exocomets, the existence of ISOs was merely hypothesised in {\it Comets II}, although their discovery was widely anticipated.  Detection of ISOs passing through the Solar system would in principle allow remote (and eventually in-situ) sampling of bodies from other planetary systems. The surety that such a discovery would eventually occur grew} with increasingly refined models of planetary system evolution, and a growing understanding of how cometary bodies are lost to interstellar space and potential evolutionary processes, outlined in \S\ref{sec:ejection}. 

Increasing observational capabilities of wide-field surveys led to the first discovery \textcolor{black}{of an ISO, 1I/`Oumuamua,} in 2017 (described in \S\ref{sec:oumuamua}). That said, the second ISO \textcolor{black}{discovery, 2I/Borisov,} was by an individual effort with a smaller telescope (see \S\ref{sec:borisov}). \textcolor{black}{2I/Borisov was clearly an active comet from first detection, matching the expectation that most ISOs would be exocometary in nature. But \Ou~was very different, with initial observations failing to detect any sign of activity, even though it was observed at a heliocentric distance of 1.2--2.8~au.} Hence, it is already clear that ISO properties span a wider range than previously thought, to the extent that it is difficult to predict what will be found in the next decade. 

In this chapter we will discuss the formation of and the observational evidence for exocomets, their ejection mechanisms into space to become ISOs and the effects that this has on their physical properties, including processing in the interstellar medium. We then describe the discovery and characterization of the first two ISOs \textcolor{black}{observed passing through} the solar system and the next decade of ISO science.
 
\section{\textbf{PLANETARY SYSTEM EVOLUTION}}
\label{sec:planetform}

\subsection{Star and primordial disk formation}
\label{sec:planetform:stars}

ISOs travel through the interstellar medium (ISM), which is the primary galactic repository from which stars, including their surrounding planetary system, form.  Since the advent of 1I/`Oumuamua, it has become clear that the ISM contains, apart from gas and dust, a third component -- interstellar objects. \textcolor{black}{These are much smaller than the free-floating planets that have been detected via microlensing \citep{Mroz:2017} and directly imaged \citep{Miret-Roig:2022}, but are much more numerous by many orders or magnitude.} Thus ISOs are present during the entire process described in the following text. 

The star and planet formation process starts in the ISM when molecular clouds form through turbulent compression and global instabilities.  Stars begin to form when denser parts of the clouds become unstable and start to collapse gravitationally. 
Once a molecular cloud begins to collapse, its central density increases considerably, eventually leading to star formation, with often an entire star cluster emerging \citep{Bate:2003}. 
A large fraction of the formed stars are not single but binaries. The binary fraction is \mbox{30--50\%} for Solar-type stars in the local neighbourhood \citep{Raghavan:2010} and up to $\approx$70\% in young clusters \citep[e.g.][]{Raghavan:2010,Duchene:2013}.
This fact might be important because some of the suggested ISO ejection mechanisms rely on binaries being present (see \S\ref{sec:ejection:individual}). 
In this phase, a nascent individual protostar grows in mass via accretion from the infalling envelope until the available gas reservoir is exhausted or stellar feedback effects become important and remove the parental cocoon \citep[for details, see ][]{Klessen:2016} and Bergin {\it et a.} in this volume.

As a natural consequence of angular momentum conservation, a disk develops during the formation of the protostar. Any initial nudge that imparts some core rotation means that material collapsing from its outer regions (with higher angular momentum) is channelled onto a disk, rather than the protostar itself \citep{Tereby:1984}.
These young stellar objects (YSOs) consist of several components: the accreting protostar, the cloud (core) it is forming from, a disk surrounding it, and outflows of material that fail to be accreted either to the star or the disk. Observationally, YSOs are classified based on their continuum dust spectrum, thus the developmental stage of its disk. Class 1 objects are still deeply embedded within the cloud and mainly emit at mm and sub-mm wavelengths, leading to a characteristic positive IR spectral index. Class 2 objects reveal sources with NIR and MIR excess, and Class 3 objects show only slight NIR excess. This classification scheme mainly reflects the development of the disks surrounding the stars from disk formation up to disk dispersal after several Myr.

\subsection{Disk properties and their development}
\label{sec:planetform:discs}

Flattened disks of cold dust and gas rotate around almost all low-mass stars shortly after their birth \citep{Williams:2011}.  These are the sites where planetesimals form. ISOs are generally believed to be ejected planetesimals. Therefore, we review the planetesimal formation process in order to understand better the properties of ISOs. 

Observations mainly at IR to millimeter wavelengths determine the frequency of disks and their mass, size, structure, and composition as a function of age. The ages of individual young stars can only be determined with a relatively large error margin. Therefore, the temporal development of disks is usually determined in young stellar clusters where many relatively coeval stars are located in close proximity. The mean age of the cluster stars is taken to test the temporal evolution of the disks in this environment.

Disk masses seem to decrease rapidly with system age \citep{Ansdell:2016}. In the extremely early phases, the disk mass can be comparable to the mass of the protostar, but after just 1 Myr, it is often less than 0.1\% of the star's mass. Some material is lost through outflows; some matter accretes onto the star or the star's radiation photo-evaporates it. Significantly in this context, some material also condenses into centimeter- or larger-sized bodies, including planetesimals. It also seems that disk dust masses correlate with stellar mass ($R_{disk} \propto M_*^{0.9}$), but this correlation steepens with time, with a more substantial drop for low-mass stars \citep{Pascucci:2016,Ansdell:2016}. Observations show that a surprisingly large fraction of disk dust masses appears low compared to the solid content of observed exoplanets \citep{Najita:2014}. This has been interpreted as a sign of early planet formation \citep{Manara:2018,Tychoniec:2020}. However, recent work suggests that detection biases might play a role here and that disks contain similar amounts of solids as found in exoplanetary systems \citep{Mulders:2021}.

Observations suggest a shallower density profile in the inner disk ($\Sigma \propto r^{-1}$) and steeper at larger distances  \mbox{($\Sigma \propto r^{-3}$)} \citep{Andrews:2020}. However, the classical assumption of a smooth gas disk is not always appropriate. High-resolution observations with ALMA provide evidence for rings, gaps and spiral structures in at least some of the most massive disks. Pronounced gap structures are often interpreted as being carved by protoplanets, implying that planetesimal formation starts early in these environments. The current sample of very high-resolution measurements in the mm continuum or scattered light is biased in favor of larger, brighter disks that preferentially orbit more massive host stars \citep{Andrews:2018, Garufi:2018}. Therefore, any conclusions about the prevalence of substructure have to be carefully assessed. 

The typical disk sizes in dust measured by mm continuum observations lie in the range of 10 -- 500 au. Gas disk sizes measured from the CO line emission are often larger than the dust disk size of the same disk.  The disk dust radius appears to be correlated with the disk dust mass \citep{Andrews:2020,Sanchis:2021}.  

The disk lifetime is not only of prime importance for planet formation theory but equally for determining the timescale on which planetesimals form. The mean disk lifetime is mostly determined from the observed disk frequency of clusters of different mean ages. It is found that the disk fraction decreases rapidly with cluster age with $<$10\% of cluster stars retaining their disks for longer than 2-6 Myr \citep{Haisch:2001,Mamajek:2009,Richert:2018} in solar neighborhood clusters. However, recently several disks have been found with ages $>$ 10 Myr, some as old as 40 Myr that still have the potential of forming planets \citep{Flaherty:2019}. Besides, the disk lifetime seems to be much shorter in dense clusters and longer in co-moving groups. The method of using disk fraction in clusters for determining disk lifetimes has been demonstrated to suffer from several biases \citep{Pfalzner:2022}. Looking at the transition phase from Class II to Class III objects, \citet{Michel:2021} found a higher typical disk lifetime of \mbox{$\approx$ 8 Myr}. A more complex picture is emerging, where disk lifetimes can range from $<$1 Myr to a few tens of Myr, and the mean age of the field star population lies in the range 1--10 Myr, but this is still not definitely determined. 

\begin{figure*}[ht!]
\begin{center}
\includegraphics[width=17cm]{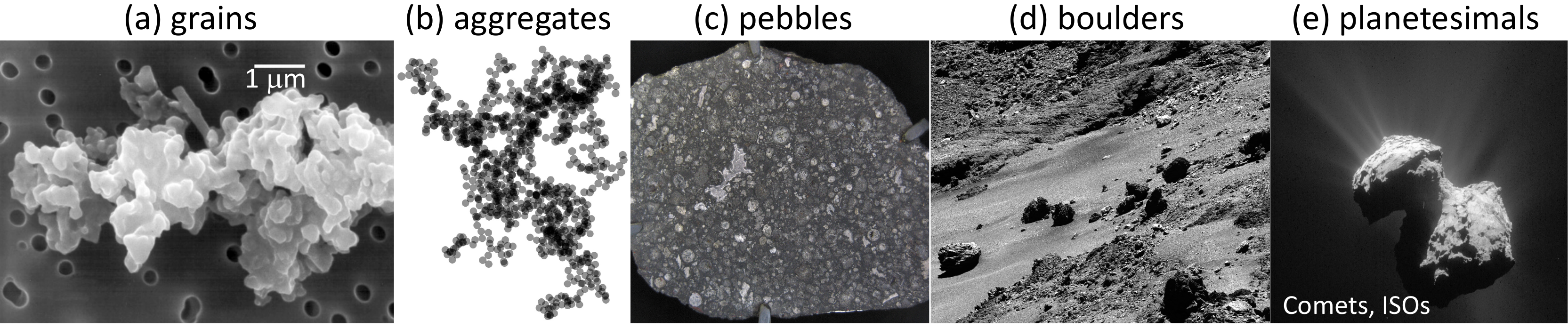}
\caption{Size scales during dust growth. Images from: (a) Chondritic interplanetary dust particle from \cite{Jessberger:2001} (CC license Attribution 2.5); (b) dust aggregate simulation from \cite{Blum:2022}; (c) 4.5 Gy old Allende meteorite with chondrules (CC Generic License 2.0); (d) images of boulders seen on by {\it Rosetta} comet 67P on 16 May 2016 from a distance of 8.9 km (ESA/Rosetta/NAVCAM - CC BY-SA IGO 3.0) (e) nucleus of 67P seen by Rosetta on 7 July 2015 from a distance of 154 km with a resolution of 13.1 m/pixel (ESA/Rosetta/NAVCAM - CC BY-SA IGO 3.0). 
}
\label{fig:dust_growth}
\end{center}
\end{figure*}

\subsection{Dust and planetesimal growth}
\label{sec:planetform:dust_growth}

For planets that reside close to their star, dust accretion is the standard formation scenario. Here planet formation takes place over a variety of size scales (see Fig. \ref{fig:dust_growth}). Initially, the mm-sized dust particles largely follow the movement of the gas. However, Brownian motion means that collisions between the dust particles take place nevertheless. The low relative velocity of the dust particles means that they stick upon contact so that first larger porous fractal aggregates form. These compact in subsequent collisions and settle toward the midplane \citep{Dubrulle:1995}, where the growth sequence continues. Once particles reach a size where they start to decouple from the gas aerodynamically, these solids start to migrate radially inwards. Further collisions lead to the formation of boulders, and eventually, planetesimals emerge. The largest planetesimals accumulate nearby smaller ones, grow further, and become terrestrial planets or the cores of gas giants. However, it is the planetesimals that are of particular interest in this chapter. Today's comets and interstellar objects are thought to be largely unchanged ancient planetesimals \citep[for details, see, for example][and Simon {\it et al.} of this book]{Blum:2008,Morbi:2016,Armitage:2018}.

This model has two potential problems: growth barriers during the accretion process and a relatively long formation timescale. The various growth barriers are the bouncing, fragmentation and drift barriers. The times expected for the different growth phases are $10^2-10^4$~yr to form mm- to cm-sized pebbles, $10^4-10^6$~yr until the planetesimal stage is reached, $10^6-10^7$~yr to form terrestrial-type planets, and an additional $10^5$~yr for the gas giants to accumulate their gas \citep{Pollack:1996}. The cumulative time seems at odds with observations of the disk frequency in young clusters, which indicate the median protoplanetary disk lifetime to be merely 1--3 Myr for both dust and gas \citep{Haisch:2001}. 

However, over the last decade, mechanisms have been devised that might solve the timescale problem while at the same time overcoming the growth barriers \citep{Armitage:2018}. These mechanisms are the streaming instability \citep{Youdin:2005} and pebble accretion \citep{Lambrecht:2012}.

In the outer parts of the disks, planets also might form via a second process. In the initial phases, these disks are still very massive and can become gravitationally unstable. Simulations show that such disk instabilities lead to spiral arm formation, which can fragment directly to form large protoplanets \textcolor{black}{\citep{Boss:1997}}. However, the material near the star stays too hot to go unstable, so this process is expected to generate planets typically only at large distances ($\geq$ 100 au) from the star. Planetesimals would be produced as a by-product of planet formation. \textcolor{black}{Details of the growth process including relevant references are described in the chapter on planetesimal formation by Simon {\it et al.} in this volume.}

\subsection{Outer areas of young disks}
\label{sec:planetform:outerdisc}

In \S\ref{sec:planetform:discs}, we saw that the gas and dust disk differ quite often in size. The difference in size is a result of the grain growth in the disk. Initially, the pressure gradient of a disk exerts an additional force that causes gas to orbit at a slightly subkeplerian speed. As long as the dust is coupled to the disk, it just follows the gas movement. However, as dust grains grow to about mm sizes, the orbiting grains experience a frictional force, drifting inwards. At the same time, the disks' viscosity means that the gas spreads out to conserve angular momentum and enable gas close in to accrete onto the star. Observations at (sub-)mm wavelengths typically trace the large dust grains; thus, disks appear more extended in gas than dust. The combined effects of growth and vertical and radial migration of dust particles mean that the disks' mean particle size should decrease with distance to the midplane and the central star. Alternatively, in terms of timescales, dust growth should proceed slower off the midplane and in the outer disk areas \citep{Dullemond:2004,Birnstiel:2014}. The question is where do the planetesimals that are ejected and become ISO primarily form? 

In the outer disk areas, \textcolor{black}{at particular distances from the central protostar, the temperature is so low that volatile compounds such as water, ammonia, methane, carbon dioxide, and carbon monoxide can condense into solid ice grains. These distances are referblack to as the snowlines. These snowlines} are essential in the context of dust growth. Ices can change the effective particle strengths and thereby affect collision outcomes.  \citet{Pinilla:2017} argue that the critical velocity for fragmentation increases at the (CO or) CO$_2$ and NH$_3$ snowlines and decreases at the H$_2$O snowline.

However, direct imaging has observed exoplanets at large distances ($>$ several tens of au) from their parent stars. Either these planets have formed there or migrated outwards to these orbits.  Among these directly imaged planet hosts, there are several that are thought to be relatively young (\mbox{$<$ 20 Myr;} for example, PDS 70 \citep{Mesa:2019}), which illustrates that even in the outer disks, planetesimals must be able to grow relatively fast. If protoplanets can grow on such a timescale, even in the outer disk, planetesimals and the resulting ISOs should form on even shorter timescales. 

\section{\textbf{EXOCOMETS}}
\label{sec:disk}

\begin{figure*}[ht!]
\begin{center}
\includegraphics[width=15cm]{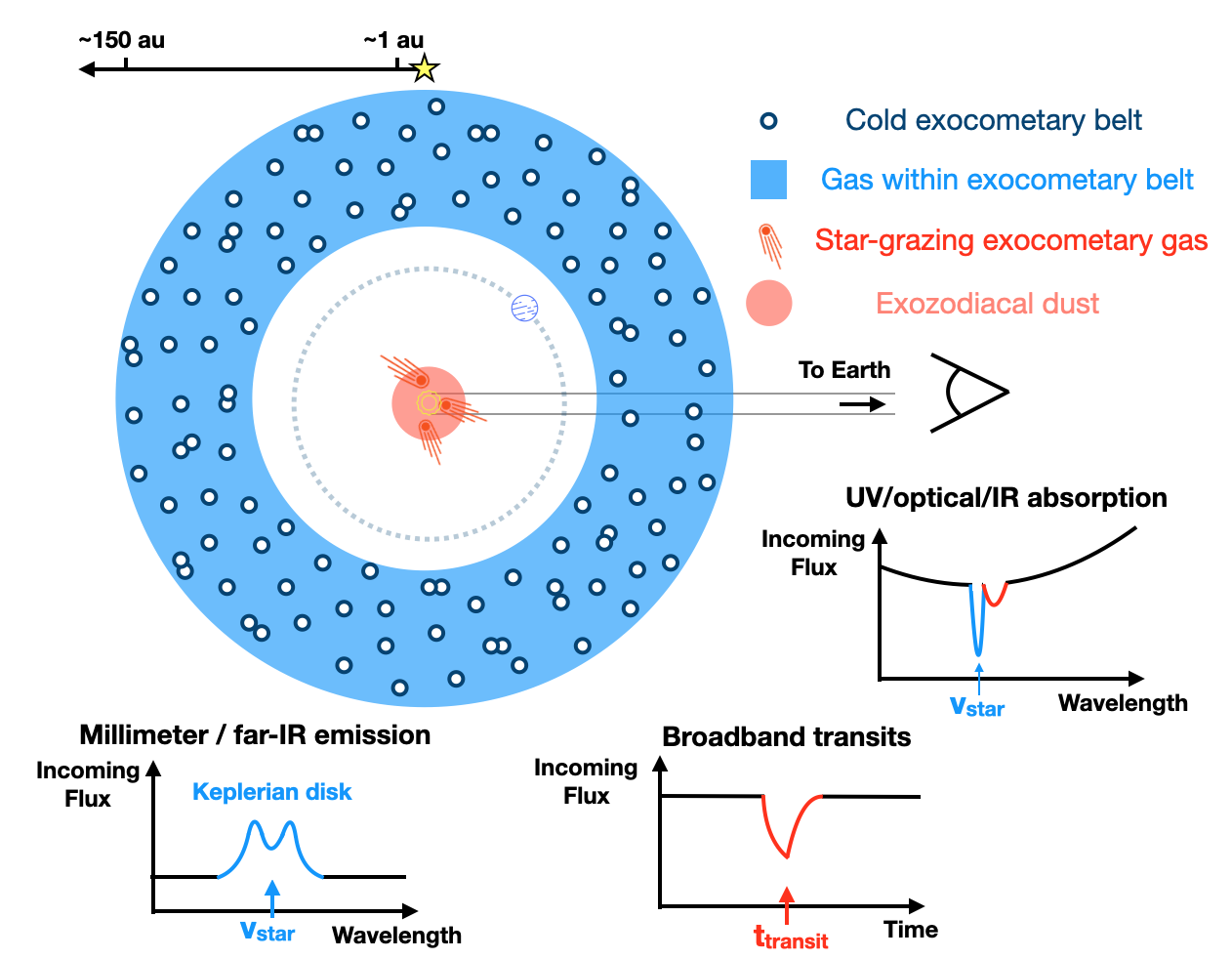}
\caption{Top-down view of a young, $\sim10-100$ Myr-old planetary system after dissipation of the protoplanetary disk. The system has largely formed outer gas and ice giant planet(s) \textcolor{black}{(at few to tens of au, like the Neptune analog with its orbit depicted as the dashed line)}, has an exocometary belt at tens of au producing cold gas and dust, and star-grazing exocomets in the inner region, producing gas and potentially warm exozodiacal dust as they approach and recede from the star. Exocomets can be probed in a variety of ways. In the outer regions (blue/cyan symbols), cold exocomets (extrasolar KBO analogs) can be probed in the belts they formed in, through cold dust and gas emission, and cold absorption against the star for edge-on systems. In the inner regions (black symbols), exocomets are probed through warm exozodiacal dust emission, as well as blue- and black-shifted absorption (see also Fig. \ref{fig:betapic:comets}) and asymmetric 'shark-fin' transits for edge-on systems.}
\label{fig:exocom_cartoon}
\end{center}
\end{figure*}

Exocomets are small bodies orbiting other stars that exhibit signs of activity, through the release of dust and/or gas \citep{Strom:2020}. As such, their definition is broader compablack to the solar system population, which is much more refined dynamically and compositionally. Evidence for exocomets dates back to 1980s, when exocomets were inferblack from variable absorption features in the spectrum of $\beta$ Pic \citep{Ferlet:1987}, and when the dust from an exocometary belt was first discoveblack around Vega \citep{Aumann:1984}. In general, exocometary release of gas and dust is detected \textcolor{black}{in both the inner (few stellar radii to few au) and the outer ($\sim10$ to $100$s au) regions} of extrasolar planetary systems (as depicted in Fig. \ref{fig:exocom_cartoon}), and we will distinguish between these two populations in the following sections. 

\subsection{Exocometary belts: extrasolar cometary reservoirs}
\label{sec:exocometarybelts}

\subsubsection{Detection and basic observational properties}

Exocometary belts (also known as \textit{debris disks}) are extrasolar analogs of our Kuiper belt, and are the reservoirs of icy exocomets in planetary systems. These belts are first detected through the dust produced by exocomets as they collide and grind down to release small, observable dust. Thermal emission as well as scattered starlight by this dust can be detected and used to study these belts. The majority of detectable belts lie at distances of tens of au from their host star, and are cold with typical temperatures of tens of K \citep{Chen:2014}. They are therefore first detected by mid- to far-IR surveys, most sensitive to the peak of the dust's thermal emission, with space telescopes such as IRAS, \textit{Spitzer}, WISE, and \textit{Herschel} \citep[e.g.][]{Su:2006,Sibthorpe:2018}. These surveys show belts to be ubiquitous; they are detected around $\sim$20\% of nearby, several Gyr-old field stars, even though the surveys' sensitivity at best enabled detection of belts a few times more massive than the Kuiper belt \citep{Eiroa:2013}. This makes this number very much a lower limit; studies of younger, less evolved belts at a few tens of Myr indeed show occurrences as high as $\sim$75\% \citep{Pawellek:2021}. 

\textcolor{black}{The typical observable (i.e. not the total) mass of belts, only considering solids} of sizes up to $\sim$cm in diameter, range between $\sim$0.1-25\% of an Earth mass, or about a tenth to 20 times the mass of the Moon \citep{Holland:2017}. \textcolor{black}{This is but a small fraction of the belts' total mass, which is dominated by the largest, unobservable bodies.} The \textcolor{black}{total mass of the Kuiper belt is estimated from observations of its largest bodies, and is about 6\% of an Earth mass \citep{DiRuscio:2020}. The Kuiper belt's dust mass is however predicted to be much lower than that of exocometary belts, to the point that a true Kuiper belt analog would not be detectable around nearby stars \citep{Vitense:2012}. In summary, detectable exocometary belts have dust masses typically $\sim10-10^5$ times higher than inferred for the present day Kuiper belt, where this broad range represents the spread of belt dust masses observed.}

\subsubsection{Birth of exocometary belts}

Exocometary belts are born at tens to hundreds of au from the central star within protoplanetary disks, but they cannot be identified until after the protoplanetary disk is dispersed. During dispersal, primordial gas and dust are removed leaving behind second-generation dust (and gas), produced destructively in collisions of larger bodies. The dispersal of the protoplanetary disk takes place rapidly once the star is a few to $\sim$10 Myr-old (\citealt{Ercolano:2017} and \S\ref{sec:planetform:discs}). The emergence of an exocometary belt from this dispersal is evidenced by the presence of an infrared excess, but with a decrease of $\sim$2 orders of magnitude in dust mass \citep{Holland:2017} compared to protoplanetary levels.

In practice, the largest exocomets within belts must form during the protoplanetary phase \textcolor{black}{to access the gas needed for the planetesimal formation process (see the chapter by Simon {\it et al.}, this volume)}, but observationally validating this is difficult. An interesting proposal is that exocometary belts may originate in (one or more of) the large rings observed in structured protoplanetary disks \citep{Michel:2021}, which has been shown could potentially explain the population of observed exocometary belts \citep{Najita:2022}. \textcolor{black}{Another possibility is that the photoevaporative dispersal of the gas-rich protoplanetary disk induces high dust-to-gas ratios \citep{Throop:2005} that may trigger the streaming instability and produce massive belts of planetesimals/(exo)comets \citep{Carrera:2017}. However, such massive, large belts are not observed, and other models of dust evolution in a photoevaporating disk indicate that the grains efficiently drift inwards faster than they can pile-up to trigger planetesimal formation \citep{Sellek:2020}.} 

One difficulty in the protoplanetary disk - exocometary belt \textcolor{black}{observational} comparison is that most of the protoplanetary disks in nearby star-forming regions surround low-mass stars, whereas the majority of detectable exocometary belts orbit around A-F stars, likely due to current lack of sensitivity to belts around later-type stars, particularly M dwarfs \citep{Luppe:2020}. In general, the presence of evolved Class III disks in young star forming regions, with masses consistent with young exocometary belts, suggests that exocomets can form early on, within the first $\sim$2 Myr \citep{Lovell:2021}.

A key component to the transition from a protoplanetary disk to an exocometary belt is the presence of gas. This produces drag, slowing collisions and favoring dust growth in protoplanetary disks; on the other hand, gas dispersal enables the more collisionally destructive, dust-producing environment of exocometary belts \citep{Wyatt:2015}. Unfortunately, distinguishing primordial from second generation gas is even harder, due to the surprisingly low CO abundances observed in many protoplanetary disks \citep{Krijt:2020}, combined with the presence of gas now discovered in a growing number of exocometary belts (see \S\ref{sec:gasbelts}), and the difficulty in measuring total gas masses in both evolutionary phases.

\subsubsection{Evolution of exocometary belts}

The physics of observable (massive) exocometary belts is driven by collisions, causing the grind-down of large exocomets (analogs to the solar system's observable Kuiper belt objects) into small, observable grains. This produces a collisional cascade, which rapidly reaches a quasi-steady state size distribution where the number of solid particles scales as $n(a)\propto a^q$, where $q=-3.5$ for a cascade extending to infinitely large and small bodies \citep{Dohnanyi:1969}. More detailed cascade simulations show that this $q$ value can vary between -3 and -4 \citep{Gaspar:2012, Pan:2012}. Observations of mm-cm wave spectral slopes implying values varying between $\sim-(2.9-3.8)$ for observable grains in the mm-cm size regime \citep{Norfolk:2021}, although interpretation of measurements in terms of size distributions is complex \citep{Lohne:2020}. These slopes imply that what extrasolar observations are sensitive to (the solids' cross-sectional area) is dominated by small \textcolor{black}{up to $\sim$cm-sized} grains, but the total mass is dominated by the largest, unobservable bodies. This is the opposite to observations of the Kuiper belt, where we are most sensitive to the largest bodies.

At the bottom end of the cascade, grains are sensitive to radiation forces, and are typically removed by radiation pressure from the central star \citep{Burns:1979}, or by stellar winds which are particularly important for young late-type stars  \citep{Strubbe:2006, Augereau:2006}. For a radiation pressure-dominated environment, the smallest, \textit{blow-out} grain size in the cascade is the one where the radiation pressure force dominates over gravity, pushing smaller grains on unbound orbits and rapidly ejecting them from the system \citep{Backman:1993}. 

This removal of small grains causes a net mass loss of material from the cascade as a function of time. In the simplest model of a collisional cascade, this would cause a belt's mass to initially stay constant - until the largest bodies begin colliding - and then decrease as $t^{-1}$ \citep{Dominik:2003} although models with more realistic treatment of cascade processes predict somewhat different time decays \citep{Lohne:2008, Kenyon:2017}. Nevertheless, a decrease in dust emission as a function of time is unambiguously observed in surveys of debris disks of different ages, and can largely be explained by collisional evolution models \citep{Sibthorpe:2018}. Additional mass removal effects could come into play that may speed up this mass loss, as hinted at by recent observations \citep{Pawellek:2021}. Notably, faster mass removal could be caused by gravitational interaction between planets and exocometary belts, potentially ejecting small bodies from the planetary system and contributing to the population of interstellar objects (see \S\ref{sec:ejection}). Such interaction could be akin to the late dynamical instability and/or to the outward migration of Neptune inferred to have depleted the vast majority of material in our solar system's Kuiper belt \citep{Malhotra:1993, Gomes:2005}.

\begin{figure*}[ht!]
\begin{center}
\includegraphics[width=16cm]{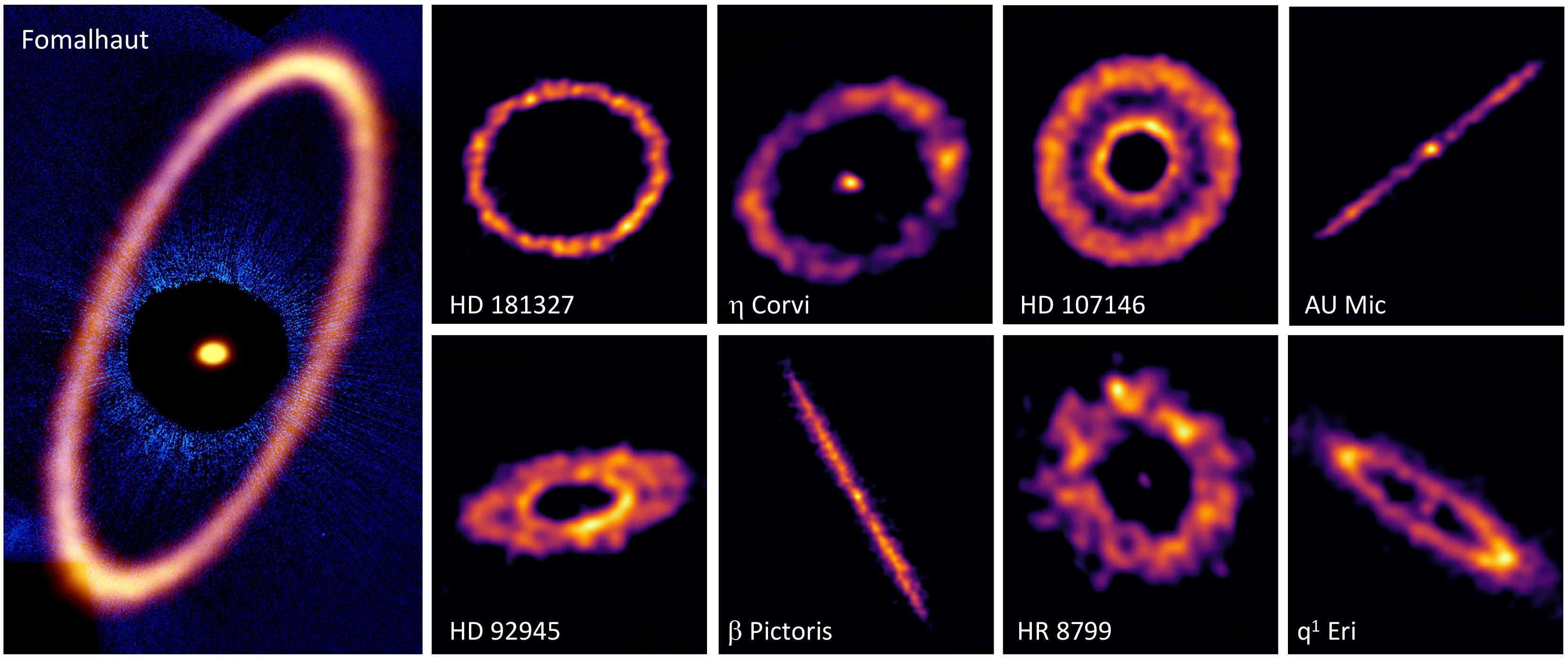}
\caption{Images of debris disks and exocometary belts. Fomalhaut credit:  ALMA (ESO/NAOJ/NRAO), M. MacGregor; NASA/ESA Hubble, P. Kalas, B. Saxton (NRAO/AUI/NSF); other images are adapted from \citet{Marino:2022}.
}
\label{fig:exocomets}
\end{center}
\end{figure*}

\subsubsection{Resolved dust imaging: structure and (exo)planet-belt interaction}
\label{sec:dust-planet}

Resolved imaging is ultimately needed to study the location and structure of exocometary belts. Given the star's luminosity, the observed temperature from the dust's unresolved spectrum can be used to calculate its location/radius (distance from the star), under the assumption that the dust behaves as a blackbody. In reality, small dust dominating the spectrum's emission is an inefficient emitter, and thus such temperature-derived \textit{blackbody radii} typically underestimate the true belt locations by a factor of a few \citep{Booth:2013}.

Belts are detected around stars from as far as nearby star-forming regions ($\sim150$ pc) to our nearest neighbors (a few pc away), with typical angular sizes of a few tenths to a few tens of arcseconds on-sky, so easily resolvable with the latest generation of high resolution facilities from optical to mm wavelengths. Given their low masses and large on-sky extents, the difficulty in imaging exocometary belts arises often from their low surface brightness, requiring deep integrations. Nevertheless, several famous belts are bright enough for structural characterization at high resolution.

The first resolved exocometary belt was that around $\beta$ Pic \citep{Smith:1984}. Almost four decades later we now have a considerable inventory of images for a few tens of belts (e.g. Fig. \ref{fig:exocomets}). These have been obtained across the wavelength spectrum, from space and ground-based optical/near-IR scattered light images \citep[milliarcsec resolution, e.g.][]{Apai:2015,Wahhaj:2016,Esposito:2020}, to space-based IR \citep[few arcsec resolution, e.g.][]{Booth:2013} and mm-wave ground-based interferometric images \citep[sub-arcsecond resolution, e.g.][]{Marino:2018a, Sepulveda:2019}.

The most basic measurable quantity of a resolved belt is its location (radius). Exocometary belts resolved across the wavelength range show typical radii of a few tens to 100s of au, with a shallow trend indicating that more luminous stars tend to host larger belts \citep{Matra:2018a,Marshall:2021}. While a combination of observational bias and radius-dependent collisional evolution can explain this positive trend, the low scatter suggests that belts (including our own Kuiper belt) may form at preferential locations in protoplanetary disks. If confirmed with a larger sample of belts, this would imply a potential connection to either temperature-dependent exocomet formation processes (e.g. the CO ice line), or to the timescale of planet formation processes preventing exocomets from growing further into planets at these distances. 

While the classic picture of an exocometary belts is that of a narrow ring, recent imaging at IR and mm wavelengths shows that broad belts (with width/radius, $\Delta R/R$, approaching $\sim1$) are at least as common as narrow rings ($\Delta R/R\ll1$; see Fig.~\ref{fig:exocomets}). Some of these belts may be born narrow but others, like the famous HR 8799 belt, may have evolved to host a scattered disk of high eccentricity particles akin to our Kuiper belt's, through gravitational interaction with interior planets \citep{Geiler:2019}. 

Beyond a simple radius and width, exocometary belts observed at sufficiently high resolution show structural features which have long been linked to the gravitational action of planets (or dwarf planets) interior, exterior, or within these belts \citep[e.g.][]{Pearce:2022}. (Sub-)structures can be broadly separated into radial, vertical, and azimuthal features.

In the radial direction, a belt's inner/outer edge and its sharpness can be linked to dynamical truncation by planets just interior or exterior to it \citep{Chiang:2009}. This is particularly true for inner edges with surface density profiles steeper than the $r^{\alpha}$ with $\alpha\sim2-2.3$ expected from collisional evolution of an undisturbed disk \citep{Schuppler:2016,Matra:2020}. 

In the absence of truncating planets, the sharpness of belt edges may also be used to probe the eccentricity dispersion, and hence the level of dynamical excitation of exocomets within the belt \citep{Marino:2021, Rafikov:2023}. Dynamical excitation of exocometary material within belts can also be probed by the belts' observed vertical structure in edge-on systems \citep{Matra:2019b, Daley:2019}. Like the Kuiper belt, this has the imprint of the system's dynamical history; for example, the vertical structure of the $\beta$ Pic belt supports the presence of a double population of exocomet inclinations \citep{Matra:2019b}, akin to the hot and cold populations of Kuiper belt objects (KBOs) in the solar system \citep{Brown:2001}.

While the radial distribution of some wide exocometary belts appears to be smooth \citep{Matra:2020, Faramaz:2021}, at least three systems so far \citep[e.g.][]{Marino:2018a} show the presence of rings separated by a gap, reminiscent of the rings observed in younger protoplanetary disks \citep{Andrews:2018}. In the simplest scenario, a single stationary planet at the gap location (tens of au) can clear its chaotic zone and carve a gap of width and depth related to the planet's mass and the age of the system \citep{Morrison:2015}. However, these gaps may be too wide and require planets to migrate \citep{Friebe:2022}, or different scenarios such as secular effects due to an eccentric planet interior to both rings \citep{Pearce:2015}.

A few narrow belts, most prominently the famous Fomalhaut \citep{Kalas:2005, MacGregor:2017} and HD202628 \citep{Krist:2012, Faramaz:2019} rings are observed to be significantly eccentric ($e\sim 0.12$). Secular perturbation by eccentric planets interior to the belts can produce and maintain eccentric belts \citep[e.g.][]{Wyatt:1999}, though the narrower than predicted width of these rings may point to alternative scenarios, such as belts being born eccentric in the protoplanetary phase of evolution \citep{Kennedy:2020}.

Azimuthal asymmetries in the brightness of exocometary belts are often detected in scattered light observations, particularly with the Hubble Space Telescope \citep[HST;][]{Schneider:2014}, though these may be simply produced by a combination of disk eccentricity, grain scattering phase functions, and viewing geometry \citep{Lee:2016}, rather than true azimuthal asymmetries in their density distribution. Similarly, eccentric disks observed in thermal emission may produce brightness asymmetries due to pericenter and/or apocenter glow \citep[e.g.][]{MacGregor:2022}, which which are expected to vary with wavelength \citep{Pan:2016, Lynch:2022}. The strongest evidence for azimuthal asymmetry arises from the edge-on $\beta$ Pic belt, which displays a strong clump in dust emission at mid-IR wavelengths \citep{Telesco:2005, Han:2023} and in gas emission \citep{Dent:2014, Cataldi:2018}, but not as prominently in scattered light or ALMA dust observations \citep{Apai:2015, Matra:2019b}. Such clumps may be produced by resonant trapping by a migrating planet at the inner edge of the belt \citep{Wyatt:2003}, akin to Neptune's migration into the Kuiper belt in the young solar system \citep{Malhotra:1993}, but also by giant impacts between planet-sized bodies \citep{Jackson:2014}.

\textcolor{black}{With the clear dynamical link between exocometary belt structure and the presence of exoplanets interacting with them, it is natural to ask whether the very existence of a belt is linked to the presence (or not) of exoplanets in the same system. For example, if the brightest debris disks form from the most massive protoplanetary disks that formed exoplanets most efficiently, one may expect a positive correlation between the presence of a belt and planets \citep{Wyatt:2007}. On the other hand, the presence of outer giant planets may efficiently eject and deplete exocometary belts (producing ISOs), while enabling terrestrial planet formation to take place interior to them without significant disruption; in this case, one would expect an anticorrelation with outer giant planets, and a correlation with low-mass planets \citep{Raymond:2011, Raymond:2012}. Evidence of trends continues to be debated in the observational literature: a correlation between low-mass planets and belts (through their infrared excess) had initially been found \citep{Wyatt:2012, Marshall:2014}, but a study with an expanded sample found no correlations with either low- or high-mass planets \citep{Moro-Martin:2015}. Similarly, no correlation was found between the presence of radial velocity planets and the properties of exocometary belts \citep{Yelverton:2020}. It is notable however that a significant number of long-period, young directly imaged planets are found in systems with bright debris disks, with some evidence of a correlation \citep{Meshkat:2017}.}

In summary, the structure of exocometary belts at tens of au probes the orbits of the icy bodies within it, and, as is the case in our solar system's Kuiper belt, can reveal dynamical interactions with mostly unseen outer planets. As proven by dynamical simulations successfully reproducing these structures, these gravitational interactions can remove exocomets in two ways. They can either eject exocomets from outer belts, potentially producing Oort clouds analogs, or directly turning them into ISOs (\S\ref{sec:ejection:planet}), but also to scatter them inwards, which could eventually produced the transiting exocomets observed close to the central stars (\S\ref{sec:innerexocomets}) and deliver volatile-rich material to planets in the habitable zones of their host stars.

\subsubsection{Evidence for gas within outer exocometary belts}
\label{sec:gasbelts}

Since their discovery almost 4 decades ago, exocometary belts have been considered to be gas free, to the point that this was long considered a defining difference between younger protoplanetary and more evolved debris disks \citep{Zuckerman:1995}. However, evidence for circumstellar gas in these systems was discovered as early as 1975 around $\beta$ Pic \citep{Slettebak:1975}, in the form of narrow Ca II H and K absorption at the stellar velocity (i.e. not significantly red- or blue-shifted, as is instead treated in \S\ref{sec:innerexocomets}) which was later spatially resolved to extend out to 100s of au from the star, where the dusty belt lies\citep{Brandeker:2004}. For a long time, evidence for cold gas in these systems remained limited to the $\beta$ Pic and 49 Ceti belts. Their near edge-on geometry allowed detection of gas in both emission and absorption against the central star \citep{Zuckerman:1995, Dent:2005, Roberge:2000, Roberge:2014}.

The evidence for exocometary gas has rapidly strengthened in the past decade. The sensitivity advance to cold gas emission brought by \textit{Herschel} and especially ALMA has led to detections of atomic (C I, O I and/or C II) and/or molecular (CO) emission within over 20 exocometary belts, largely around $\sim$10-40 Myr old stars \citep{Moor:2017}, but also as old as 440 Myr \citep[Fomalhaut,][]{Matra:2017b}. Early ALMA CO gas detections highlighted a dichotomy between belts with very high CO masses, approaching those of protoplanetary disks \citep[$10^{-2}-10^{-1}$ M$_{\oplus}$, e.g.][]{Kospal:2013}, and more tenuous CO gas disks \citep[$10^{-7}-10^{-5}$ M$_{\oplus}$, e.g.][]{Matra:2017a}. 

The origin of the gas remains an open question for high-mass CO systems. In a \textit{primordial} scenario, the gas is a remnant of the protoplanetary disk (contrary to the dust, which is second-generation, leading to the term \textit{hybrid} disk); H$_2$ dominates the gas mass, allowing CO to remain shielded from stellar and interstellar photodissociation over the lifetime of the system \citep{Kospal:2013}. In a \textit{secondary} scenario, CO gas is continuously produced by exocomets within the belt, just like the dust \citep{Zuckerman:2012, Matra:2015, Marino:2016}. However, the production rate is unlikely to be sufficient to maintain the high CO masses observed, so a shielding agent other than H$_2$ (which is only produced in trace amounts in solar system comets) is needed. Shielding CO photodissociation (producing C and O) by C photoionization is a promising scenario to produce what would be second-generation, shielded high-mass CO disks \citep{Kral:2019}.

For low-mass CO belts, like Fomalhaut, HD181327 and $\beta$ Pic, the secondary scenario is heavily favored, as the CO photodissociation timescale, even when accounting for the potential of unseen H$_2$ with primordial CO/H$_2$ ratios, is much shorter than the age of the system \citep[e.g.][]{Marino:2016}. This simple argument implies that CO must have been recently (and likely continuously) produced, and must be of secondary, exocometary origin. 

\subsubsection{Gas release processes, composition, and evolution}
\label{sec:gasrelease}

The gas release process of exocomets within belts from their host stars is likely different from solar system comets. In the very cold ($\sim$tens of K) environments at tens of au, the heating and sublimation-driven process producing gas in single solar system comets and ISOs (as they approach the Sun) is unlikely to be efficient; while activity suggesting CO outgassing takes place out to Kuiper Belt distances in at least some long period comets \citep[e.g. C/2017 K2 (Pan-STARRS)][]{Meech:2017b}, no activity is detected from KBOs on stable orbits \citep{Jewitt:2008} including Arrokoth \citep[(486958) 2014 MU69,][]{Lisse:2021}. \textcolor{black}{Care should be taken in associating locations with temperatures in the Kuiper belt versus exocometary belts. As exocometary belts are most commonly observed around more luminous, A-F type stars, they will have warmer temperatures at the same radial location. This is however mitigated by the temperature dependence with stellar luminosity being shallow for blackbody-like grains heated by starlight ($T\propto L^{0.25}$), combined with the fact that belts tend to be located at larger radii around more luminous stars (\S\ref{sec:dust-planet}).}

One possibility is that we are observing a vigorous sublimation period for exocomets immediately following the dispersal of the protoplanetary disk \citep{Steckloff:2021, Kral:2021}. In this period, hypervolatiles like CO, N$_2$ and CH$_4$ sublimate at high rates due to the heating by newly visible starlight, previously shielded by large amounts of protoplanetary dust. In this scenario, Arrokoth-sized objects ($\sim$10s of km diameter) at $\sim$45 au from the central star take 10-100 Myr for the heat to reach their interior, which could explain the higher detection rate around young exocometary belts in that age range. However, in exocometary belts a large range of sizes would have to be considered, and it remains unclear whether the production rates could be sufficiently high to explain the observed CO masses.

Other physical mechanisms have been proposed to release gas in the collisionally active environment of young exocometary belts. While typical collision velocities are unlikely to result in sufficient heating, collisional vaporization of very small grains accelerated by radiation pressure could lead to ice release into the gas \citep{Czechowski:2007}. Additionally, UV photodesorption could very effectively remove volatiles such as H$_2$O from the surface of icy grains/objects \citep{Grigorieva:2007}, given these are continuously produced/resurfaced through the collisional cascade. 

Once released, in the absence of shielding, molecular gas is expected to be rapidly destroyed by the stellar and interstellar UV fields. CO (with N$_2$) is the most photo-resistant molecule, which may explain why it is the only one detected so far \citep{Matra:2018b}. The CO ice mass fraction of exocomets can then be estimated from the ALMA mass measurements and steady state production/photodestruction arguments, indicating an exocometary CO content within an order of magnitude of solar system comets \citep{Matra:2017b, Matra:2019a} - showing that accessing exocometary compositions similar to solar system comets is now possible. Unfortunately, the short photodissociation timescales and/or weak transitions of other molecules make other molecular detections currently challenging \citep{Matra:2018b,Klusmeyer:2021}.

On the other hand, atomic gas is abundant and detected both in emission and absorption for edge-on belts. The detection of atomic N  \citep{Wilson:2019} and S, as well as C and O \citep[see][and references therein]{Roberge:2006} in the $\beta$ Pic belt suggest that molecules other than CO are being released, arguing against the gas release being limited to hypervolatile molecules.  Atomic gas accumulating over time at the belt location is expected to produce a disk that viscously spreads over time, at a rate determined by the poorly-constrained gas viscosity, and eventually form an accretion disk \citep{Kral:2016}. This model can largely explain the population of CO masses observed \citep{Marino:2020}, including high CO mass disks through C shielding. However, resolved observations of C I gas by ALMA show inner holes inconsistent with accretion disk profiles, suggesting a more complex evolution \citep{Cataldi:2018, Cataldi:2020}.

\subsubsection{Detectability of exo-Oort clouds}
\label{sec:exo-oorts}

Extrasolar Oort cloud analogs may also exist, and exocomets in those stars would be heated by absorption of radiation from their central star, nearby stars and the \textcolor{black}{Cosmic Microwave Background (CMB)}. The challenge for detecting extrasolar Oort clouds is their low surface density, low temperature and large sky areas, which is why long-wavelength surveys around nearby stars are best suited for detection. An additional factor is whether dust might be produced in Oort clouds, and if so how long it might survive. \cite{Stern:1991} used {\it IRAS} to search for excess far-IR emission around 17 nearby stars but did not achieve any detections. More recently, \cite{Baxter:2018} used {\it Planck} survey data to average the sub-mm emission around selected stars within 300~pc, and also performed a directed search around Fomalhaut and Vega, but again did not detect any signal. Intriguingly, when stacking 43 hot stars within 40--80 pc, they detected a statistically significant excess flux at radial distances of $10^4-10^5$~au. \textcolor{black}{Searching around early-type stars is a key factor, as this will increase stellar heating of the exo-Oort body surfaces and aid detection}.
But they cautioned that this could be a false positive due to the fine structure of galactic dust on scales of the {\it Planck} beam width. Alternatively it could be a real signal caused by weak nebular emission or dust grains ejected from debris disks near the stars. \textcolor{black}{Future CMB  missions such as LiteBIRD \citep{Hazumi:2020} may approach the sensitivity required for unambiguous detection. However, the $0.5^\circ$ mission resolution would restrict it to searches around the nearest stars within 10~pc, for which $10^4$ au spans $\sim 0.3^\circ$. Unresolved emission from more distant stars would be ambiguous as it could also be caused by thermal emission from dust closer to the star, see \S\ref{sec:dust-planet}. }

\subsection{Exocomets in inner planetary systems}
\label{sec:innerexocomets}

\subsubsection{The $\beta$ Pictoris system}
\label{sec:BPic}

\begin{figure}[ht!]
\includegraphics[width=9cm]{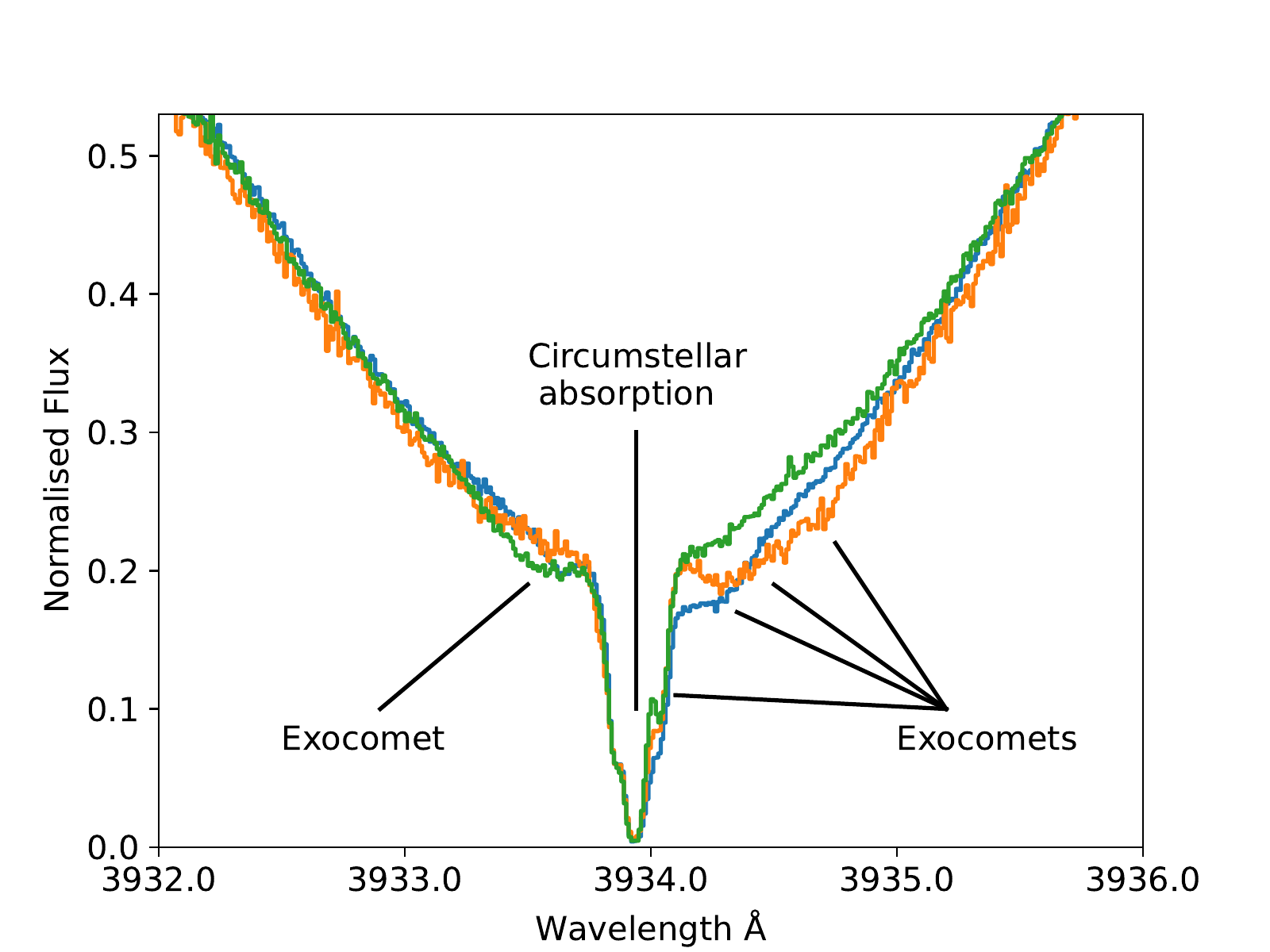}
\caption{Spectra of the core of the Ca~II K stellar line in $\beta$ Pic at three separate epochs over 4 days. The stable central narrow absorption is due to gas in the circumstellar disk, marking the stellar radial velocity. Transient Ca~II absorption features caused by individual exocomets can be seen both blue-shifted and red-shifted relative to the star (see \S\ref{sec:exocometarybelts}, and Fig.~\ref{fig:exocom_cartoon}). \textcolor{black}{Data from the ESO 3.6-m+HARPS public archive (see acknowledgments).}
}
\label{fig:betapic:comets}
\end{figure}

The star $\beta$ Pic is a nearby ($d=19.3$ pc) A6V star that was the first to have a circumstellar debris disk optically imaged \citep{Smith:1984}. Soon after, transient red-shifted absorption features (see Fig.~\ref{fig:betapic:comets}) in the UV spectrum were detected in the wings of the stellar Al~III, Mg~II and Fe~II lines \citep{Lagrange-Henri:1988}, and in numerous high-resolution optical spectra that include the strong Ca~II H\&K lines \citep{Ferlet:1987}. These were interpreted as originating from planetesimals undergoing sublimation when within a few stellar radii from $\beta$ Pic, with most features being redshifted from motion towards the star. They were subsequently termed `Falling Evaporating Bodies', or FEBs. In modern parlance these are exocomets, but care must be taken. The observation of metal lines may imply sublimation of refractory elements, and hence not necessarily a volatile-rich body as is common in our solar system. Subsequent studies of these exocomets have tended to use the strong Ca~II H \& K lines in high-resolution optical spectra, although Na~I D-line absorption is also seen. \textcolor{black}{Unfortunately only metal lines in these exocomets have been measured and the lack of detections of the volatile species seen in solar system comets has prevented a direct comparison.} While studies of the gas in the outer circumstellar disk show it is likely to be released by exocomets in those regions as discussed in \S\ref{sec:gasrelease}, there is as yet no direct link to the exocomets seen close to the star.

Akin to solar system comets, these exocomets are clearly undergoing significant mass loss, which argues for a source region in the system. \cite{Beust:2000} and \cite{Thebault:2001} had previously shown that a Jupiter-sized planet within 20~au could sustain such a population by strong orbital evolution via the 4:1 and/or 3:1 mean-motion resonances. This would place the source region at 4--5 au from $\beta$ Pic, within the imaged debris disk. The  discovery by direct imaging of the Jovian exoplanet $\beta$ Pic b by \cite{Lagrange:2009}  orbiting at $a=9.9$~au gave credence to this hypothesis (see \S\ref{sec:dust-planet} above). The more recent discovery of the closer massive planet $\beta$ Pic c at $a=2.7$~au \citep{Lagrange:2019} will result in more complex dynamical evolution, but clearly enhances the possibility of scattering exocomets existing in the observed disk into star-approaching orbits. 

Although exocomets typically transit the star over a couple of hours, \cite{Kennedy:2018} was able to measure radial acceleration by individual bodies, and constrain  their orbital parameters via MCMC methods. A larger analysis was made by \cite{Kiefer:2014} of 252 individual exocomets seen in $\sim 6,000$ spectra obtained in 2003--2011. The velocities and absorption strengths display a bimodal distribution, implying at least two dynamical populations. Exocomets creating shallow absorption lines tend to have small total absorption and higher velocities, with periastron distances $\simeq 0.08$~au (9 stellar radii) over a wide range of longitudes.  The less numerous exocomets that produce shallow absorption also exhibit smaller radial velocities due to larger periastron distances of $\simeq 0.16$~au (18 stellar radii). A small range of longitudes of periastron imply these less active objects all share similar orbits, and may result from the breakup of a single progenitor similar to the sungrazing Kreutz family
(\citealt{Jones:2018} and references therein). 

Exocomets orbiting $\beta$ Pic have now also been detected via broadband optical light curves using the Transiting Exoplanet Survey Satellite (TESS) mission \citep{Zieba:2019}. The transit signals from individual exocomets show a `shark fin' appearance caused by the asymmetric morphology of dust comae and tails, as predicted by \cite{desEtangs:1999}. Analysing all TESS light curves at that time, \cite{desEtangs:2022} identified 30 individual transits. Assuming that each exocomet on average has the same fractional surface area sublimating, the differential size distribution has a power-law exponent of $-3.6\pm 0.8$, similar to that seen in solar system comets. 

\subsubsection{Other systems exhibiting exocomet signatures}

As explored in Section~\ref{sec:planetform}, the decreasing mass of circumstellar material as a young star begins its main-sequence lifetime implies that exocomets may also be more abundant at early times. Indeed, transient spectroscopic absorption features similar to $\beta $ Pic are seen around many Herbig Ae/Be stars \citep{Grady:1996}. With $\beta$ Pic as the archetype main-sequence host star, exocomets have been confirmed via similar spectroscopic or photometric transient features around 3 other stars: 49 Cet, HD 172555 and KIC 3542116 (\citealt{Strom:2020} and references therein). In addition there are likely exocomet systems such as c~Aql (HD~183324), which exhibits Ca~II variability on timescales of minutes \citep{Montgomery:2017, Iglesias:2018}. In total, spectroscopic and photometric signatures indicating probable exocomet activity have been reported for another 30 or more stars \citep{Strom:2020}.

The common factor in all of these stars is that they are early-type A-F stars with bright apparent magnitudes. The brightness allows high signal-to-noise photometry from space-based missions such as Kepler and TESS to reveal weak transit signals from stars with less stellar activity than later K/M-type stars, while B/A stars also exhibit relatively clean optical spectra with fewer stellar absorption lines,  facilitating spectroscopic detection. In terms of total numbers, a directed spectroscopic survey of 117 B--G stars by \cite{Rebollido:2020} found $\sim15$\% exhibited transient Ca~II and Na~I absorption. In summary, there is strong evidence that exocomets are commonly associated with early-type B/A stars. The evidence for later-type stars is less extensive, given that a search of 200,000 {\it Kepler} light curves by \cite{Kennedy:2019} only found 5 potential photometric transits occurring for 3 stars.

\subsubsection{Warm exozodiacal dust}

Aside from the more distant cometary belts described in \S\ref{sec:exocometarybelts}, approximately 20\% of A--K stars exhibit thermal signatures of warm ($T\sim 300$ K) dust within a few au (see \citealt{Kral:2018}). The combination of the Poynting-Robertson effect on large dust grains and radiation pressure on small dust grains implies very short lifetimes without continuous resupply. Potential sources of this dust are either asteroidal collisions or dust ejection from comets.  This may especially be the case for systems with colder outer dust disks, where the source of the dust is expected to be ice-rich bodies. Given that the generation of the solar system's zodiacal cloud is dominated by short-period comets \citep{Rigley:2022}, it is plausible that this may be also the case for some exosystems \citep{Marboeuf:2016}. A survey of exozodiacal dust around 38 stars showed a significant correlation with the existence  of cold dust disks but no correlations with stellar age, as would be expected by analogy with the solar system \citep{Ertel:2020}.

An exemplar star where this appears to be the case is the main-sequence F2 star $\eta$ Corvi. This star possesses a cold outer belt centered at $\sim 165$~au, plus a warm inner dust ring whose inner boundary is as close as 2--3~au (\citealt{Duchene:2014} and references therein). Near and mid-IR spectra were modelled by \cite{Lisse:2012} to reveal large quantities of amorphous carbon and water ice-rich dust in the inner ring, as well as silicates. They interpret this as the recent break-up of a large centaur-like icy body due to a collision with an even larger silicate body, resulting in most of the current observed warm dust. Inwards perturbation of icy bodies from the outer cold belt by undetected planets as simulated by \cite{Bonsor:2014} is supported by the large gap between the inner and outer disks, plus the subsequent detection of CO at $\sim 20$~au which could be due to inwardly evolving exocomets undergoing CO sublimation \citep{Marino:2017}. 

\subsection{Late stages of stellar evolution}

\subsubsection{Exocomets during the late stages of stellar evolution}

Once a star evolves off the main-sequence, the large changes in luminosity in the Red Giant/Asympototic Giant Branch (RG/AGB) stages, together with the mass-loss at the end of the AGB phase, will result in significant physical and dynamical changes to any orbiting cometary bodies \citep{Veras:2016}. For example, during the Sun's RG phase the water sublimation boundary will move from $\sim 3$~au to $\sim 230$~au, leading to cometary activity throughout the Trans-Neptunian region \citep{Stern:1990a, Stern:2003}. 

\cite{Melnick:2001} used the Submillimeter Wave Astronomy Satellite ({\it SWAS}) to discover a large amount of water vapor around the carbon-rich AGB star IRC+10216. This was initially interpreted as originating by ongoing sublimation of exocomets in that star's Kuiper Belt. The inferred mass of cometary bodies was $\sim 10_\oplus$ \citep{Saavik:2001}. However, a later survey of eight AGB stars with {\it Herschel} \citep{Neufield:2011} showed a cometary source was unlikely, as the water line profiles matched that of the circumstellar CO, with widths as expected from the circumstellar expansion velocity. If the water originated from Kuiper belts, then at least some would be viewed near face-on, and the line widths of cometary-produced water would be much narrower. Instead, water formation appears to be either by shock-induced dissociation of CO, where the oxygen subsequently combines with hydrogen to produce H$_2$O, or due to UV photodissociation of CO and SiO to again produce oxygen and water vapor \citep{Lombaert:2016}. Therefore, there currently exist no strong indicators of exocomets around stars at the RG/AGB stage. However, we know they are present due to observations of stars at the very end of their stellar evolution, White Dwarfs.

\subsubsection{Evidence for small bodies around white dwarfs}

There is now undisputed evidence that planetary systems can (in some form) survive post-main sequence evolution. The origin of heavy element absorption lines in White Dwarf (WD) photospheres is now accepted as being caused by pollution from infalling planetary material. The paradigm is that small bodies near the WD undergo tidal disruption to form a debris ring, where subsequent collisions create dust that is slowly removed via the Pointing-Robertson effect to ``pollute'' the stellar photosphere.  Strong evidence for this is found in the association between WD infrared excesses due to circumstellar dust and atmospheric pollutants (see \citealt{Farihi:2016} and references therein). 

The relatively short timescales for this material to sink below the observed photosphere means this process is currently occurring where observed. The simplicity of WD photospheres results in high-precision elemental abundances being derived via model atmosphere fitting. There is no doubt that almost all small bodies providing this material were volatile-poor \citep{Gansicke:2012, Xu:2019}. However, two WDs have been identified as being polluted by water-rich bodies \citep{Farihi:2013, Raddi:2015}. This dominance of silicate-rich small bodies may not reflect the original small-body population that survives into the WD phase. A theoretical study of water retention by \cite{Malamud:2016} found that larger water-rich bodies should survive the RGB and AGB stage of stellar evolution. They propose that the observed deficit is due to water being lost prior to tidal disruption or during the disk formation stage. If the WD is in a wide binary system, then dynamical evolution of exocomets from a Kuiper-belt via Kozai-Lidov resonances can result in WD pollution over a wide range of timescales of $10^8-10^{10}$ yr \citep{Stephan:2017}. Given that 25-50\% of all WDs exhibit pollution by small bodies, exocomets may therefore be relatively common around white dwarfs.

\section{\textbf{DYNAMICAL EJECTION AND EVOLUTION OF COMETARY BODIES }}
\label{sec:ejection}

\begin{figure*}[ht!]
\begin{center}
\includegraphics[width=17cm]{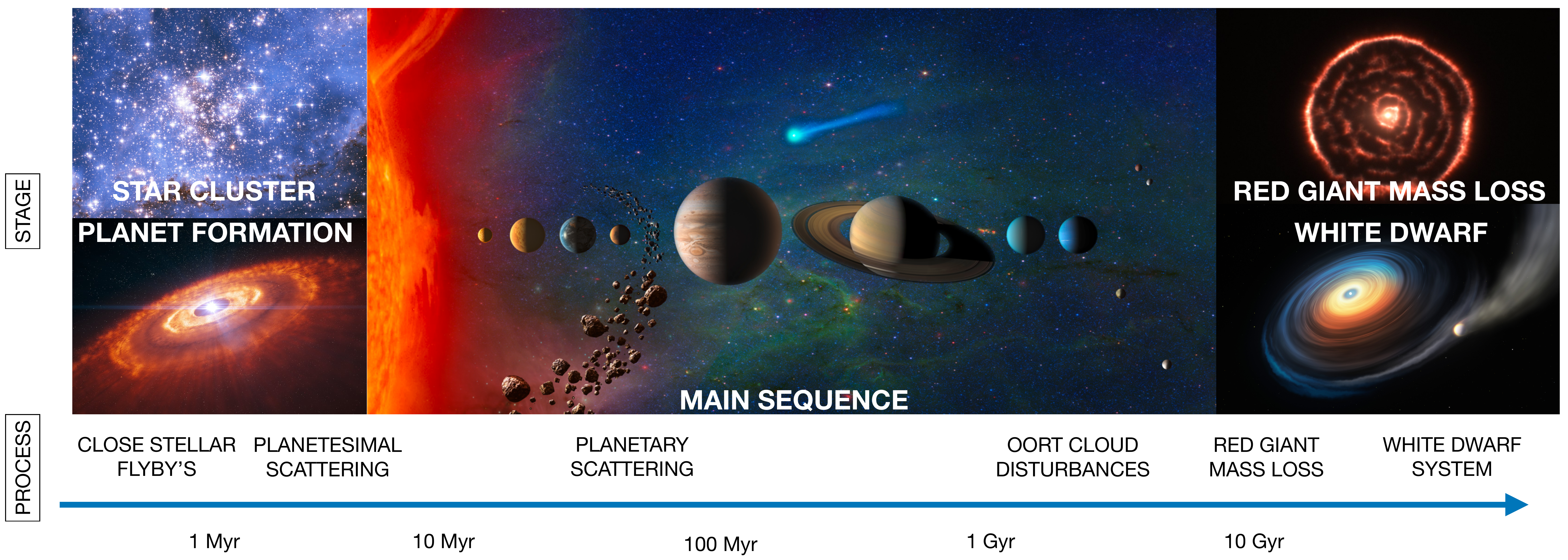}
\caption{Schematic picture of various planetesimal ejection mechanisms during the different stages of stellar evolution. Credits left to right, top to bottom; star cluster NGC~346 (NASA/STScI), planet formation (ESO/L. Cal\c{c}ada, solar system (NASA), red giant R~Sculptoris (ALMA/ESO/NAOJ/NRAO/M. Maercker et al.), white dwarf planetary evolution (ESO/M. Kornmesser).}
\label{fig:stellar:ejection}
\vspace{-0.25cm}
\end{center}
\end{figure*}

The discovery of 1I/`Oumuamua and 2I/Borisov gives evidence to the hypothesis that over a star's lifetime, part of its remnant small body population becomes unbound (e.g., \citealt{Fernandez:1984, Duncan:1987, Wyatt:2008}). Thus ISO production seems to be a natural consequence of the existence of reservoirs of small bodies in planetary systems. These small bodies can become unbound from their parent system in several ways (see Fig.~\ref{fig:stellar:ejection}). In the following, we will describe the various ISO production processes in chronological order of their occurrence during a star's lifetime. We give estimates on the order of magnitude level of the total mass of ISOs released by the individual processes. However, these values are highly sensitive to the assumed (debris) disk masses. 

\subsection{Individual ejection mechanisms}
\label{sec:ejection:individual}

\subsubsection{ISO formation during disk dispersal}
\label{sec:ejection:disc}

In \S\ref{sec:planetform:discs} we saw that disk masses decrease rapidly within the first 1--3 Myr. While initially, this mass loss will be in the form of small dust particles, later on, larger particles and eventually planetesimals will contribute to this mass loss. So far, the matter that becomes unbound during this early stage has received little attention. The main challenge is the significant uncertainties concerning the timescale on which planetesimals form. Therefore, it is unclear how many planetesimals are present in disks at this stage. \textcolor{black}{Very young disks have masses comparable to the mass of their host star ($m_d \approx M_s$).}  
Considerable amounts of ISOs could, in principle, already be produced during this stage.  \textcolor{black}{The disk masses decrease to ($m_d \approx 0.01 M_s$) within just a few ten thousand years; if any planetesimals are already present during that phase, they could become released due to the lower potential caused by the gas loss. }

In recent years, several old disks ($\>>$ 10 Myr) have been discovered still containing considerable amounts of gas. These disks likely also contain a sizable planetesimal population. When they eventually disperse their gas, \textcolor{black}{the potential energy change due to the gas loss might be enough that a fraction of the planetesimals} becomes unbound from the outer disk regions. Most proposed disk dispersal mechanisms are relatively gentle. Thus the ISOs would leave their parent star at relatively low velocities ($\ll$ 0.5 m/s). The only process that would potentially lead to asteroidal ISOs ejected at higher velocity are those involving disk jets. However, as jets only exist in the very early phases of disk development \mbox{($<$1 Myr)} it is unlikely that planetesimals had already formed.

\subsubsection{Cluster environment induced planetesimal release}
\label{sec:ejection:cluster}

Planetesimals can also be released during another star's close flybys \citep{Hands:2019}. Here mostly planetesimals of the outer disk become unbound so that the ISOs produced by this process are usually icy objects. Close flybys happen most frequently during the first 10 Myr of a star's life when the star is still part of a cluster \citep[e.g.][]{adams:2006}. However, debris disks can also be affected, as their longer lifetimes can sometimes outbalance the lower frequency of such events during later stages. 

The frequency of close flybys varies enormously between different types of stellar groups. In long-lived compact open clusters, stellar flybys are much more common than in the short-lived, more extended clusters typical for the current solar neighborhood. Based on typical disk truncation radii in diverse cluster environments, \citet{Pfalzner:2021} find that clusters like the Orion nebula cluster are likely to generate the equivalent of 0.85 $\mearth$ of ISOs per star. In contrast, compact clusters like NGC 3603 can produce up to 50$\mearth$ of ISOs per star. Our solar system probably created the equivalent of \mbox{2-3 $\mearth$} of ISOs \textcolor{black}{by this process. In clusters, }ISOs leave their parent system typically with velocities in the range of 0.5--2 km/s. 

External photo-evaporation and viscous spreading can, under certain circumstances, strongly influence the gas component of discs \citep{Adams:2010, Concha:2019}. However, the decoupling of the gas and planetesimals means that these two processes release only a small number of ISOs. The velocity of these few ISOs should be much smaller than those released due to close flybys. 

\subsubsection{Early scattering by planets}
\label{sec:ejection:planet}

During the phase when a planet is still accreting, planetesimals can receive gravitational kicks from planets. Such a perturbation can lead to two outcomes: scattering and collisions. The ejection to collision rates (assuming constant density for the planet) are given as $R_e/R_c =(M_p^{4/3}a^2)(M_*^2)$, where $M_p$ and $M_*$ are masses of the star and the planet, and $a$ is the planet’s semi-major axis \citep{Cuk:2018}. At large distances, collisions become less likely, while scattering and, therefore, ejection becomes more efficient. \textcolor{black}{The reason is that the Hill radius, which controls scattering, expands linearly with the orbital radius, whereas the physical size is fixed.}

\textcolor{black}{A planet can eject a planetesimal if the Safranov number \mbox{$\frac{1}{2} v_{esc}/v_{orb} \gg$ 1,} where $v_{esc}$ is the escape velocity and $v_{orb}$ the orbital velocity of the planetesimal \citep{Raymond:2018}. While the other giant planets all have similar Safronov numbers, it is actually the structure of the system itself that makes Jupiter responsible for the vast majority of planetesimals that have been ejected from the solar system \citep{Dones:2015}. }

Most investigations \textcolor{black}{of this process} assumed solar system-like environments \citep{Raymond:2020}. The early solar system likely had a planetesimal disk with \textcolor{black}{a mass of \mbox{ $\approx$ 5--65 $\mearth$} \citep{Deienno:2017,Liu:2022}. }The estimates of how many planetesimals have been ejected range from  0.1 $\mearth$ \citep{Raymond:2018} to 20 $\mearth$  \citep{Trilling:2017} and even  40 $\mearth$  \citep{Do:2018} per star. \textcolor{black}{The differences in estimated total masses are primarily due to the unknown nature of the size distribution of ejected planetesimals.} The ejected planetesimals from these regions would leave the systems with a typical velocity of 4--8 km/s \citep{Adams:2005}, significantly higher than for the above mechanisms. The amount of planetesimals ejected by scattering depends on the structure of the planetary system. Systems of densely packed giant planets that orbit close to a common plane are most efficient in ejecting planetesimals. 

In our solar system, the ejected planetesimals would have been primarily icy because ejection is much more efficient outside the gas giants than closer in.  Planetesimal ejection from the asteroid belt will be much less \textcolor{black}{common}. It is mainly caused by gravitational perturbations by Jupiter and Saturn and the mutual perturbations amongst the largest asteroids. \textcolor{black}{It is currently an open question whether the early asteroid belt was high-mass and depleted \citep[e.g.][]{Petit:2001} or whether it was born low-mass and later on populated by dynamical processes \citep{Raymond:2017}. If it was initially of high mass,} ejections have depleted the asteroid belt by a factor of $\sim100$ \textcolor{black}{in number}, the original asteroid belt being  $\approx$ 10–20 times more massive than today \citep{O'Brien:2003}. \textcolor{black}{In this case, the total mass of ejected asteroids would amount to $\approx$ 0.004 -- 0.008 $\mearth$.}

However, the system's giant planets eject ISOs not only while still accreting gas \citep{Raymond:2017,Portegies:2018}. In compact planet configurations, interactions between the planets themselves or the planets and the planetesimal disk can trigger instabilities. The planets move positions throughout such a phase of planet-planet scattering. During such an event, the bulk of planetesimal disk would be ejected \textcolor{black}{\citep{Raymond:2010}}. The structure of today's Kuiper belt and the Oort cloud can be explained by this mechanism. 

Many exoplanetary systems are in compact configurations that might become unstable. Moreover, the solar system's giant planets were likely originally on more compact, resonant orbits \textcolor{black}{\citep{Morbi:2007,Pierens:2008}}. Originally it was proposed that such an instability happened in the solar system well after planet formation finished (500-700 Myr); however, a consensus is emerging that the instability happened early, no later than 100 Myr after the formation of the solar system \textcolor{black}{\citep{Zellner:2017, Morbi:2018}}.
\subsubsection{Drifting from Oort clouds}
\label{sec:ejection:Oort}

Even though the solar system’s Oort cloud formation still has many unknowns, we can also expect exo-Oort clouds to exist around other stars. Oort cloud bodies are icy exo-comets and possibly asteroidal objects \citep{Weissman:1999, Meech:2016} and are only weakly bound to their parent system. Being situated at a significant distance from their parent star, they are subject to external influences, like Galactic tides and stellar flybys. Consequently, some planetesimals will always drift gently away from the star’s Oort cloud over the entire main-sequence lifetime of the star \citep{Moro_Martin:2019,Portegies:2021}. Thus their escape velocity is, on average, very low, probably $<$ 0.1 m/s. \cite{Hanse:2018} estimates the Sun will lose 25--65\% of its Oort cloud's mass mainly due to stellar encounters. \textcolor{black}{Such former Oort cloud members will contribute but amount to $<$ 10\% of the total ISOs population \citep{Portegies:2021}.}

\subsubsection{White dwarf phase}
\label{sec:ejection:death}

Finally, there is an increased planetesimal ejection rate towards the end of a star’s giant branch phase \textcolor{black}{\citep{Veras:2016, Veras:2020}}. During that phase, expansion of the stellar envelope can directly engulf closely orbiting planets and tidally draw into the envelope planets which reside beyond the maximum extent of the stellar envelope. While the planetesimal disks are mainly unaffected by engulfment, stellar mass loss plays a role. During the post-main sequence evolution, all objects, planets and planetesimals alike, move outward due to orbital expansion caused by mass loss \citep{Villaver:2014}. In the solar system, planetesimals within $\approx$10$^3$ au would all double their semi-major axes \citep{Veras:2020} \textcolor{black}{and therefore be less strongly bound to their host star}. The ISOs produced during the giant branch phase gently drift away from their parent star at very low relative velocities of 0.1 – 0.2 km/s \citep{Pfalzner:2021}. In addition, mass loss also destabilizes the planetary system, resulting in the gravitational scattering of planetoids by massive perturbers. \citet{Rafikov:2018} suggests tidal disruption events of (initially bound) planetary objects by white dwarfs as an additional ISO production mechanism. Some of these objects are scattered towards the white dwarf on almost radial orbits and get tidally shredded apart. Some of the fragments that are produced are ejected into interstellar space.

The highest mass white dwarf progenitors would yield the greatest giant branch excitation and and are most efficient in the ejection of planetesimals \citep{Veras:2020}.  This study  points out that most stars in the Milky Way are less massive than the Sun, and concludes that the production of ISOs from within $40-1000$ au of the evolved star is insignificant when compared to the number created during early stellar and planetary evolution. 

\subsubsection{Binary scenarios}
\label{sec:ejection:binary}

In \S\ref{sec:planetform:stars}, we noted that binary and multiple stars are very common in the Galaxy. Moreover, binaries are also more efficient scatterers, providing more substantial dynamical perturbations than even a gas giant planet \citep{Raymond:2017, Cuk:2018, Jackson:2018}. All the processes discussed before for single stars could also take place in binary systems. So far, more or less exclusively scattering processes have been considered in detail. Close binary stars have well-defined dynamical stability limits, and any objects entering within a critical orbital radius are destabilized. \citet{Jackson:2018} assume that planetesimals form beyond this stability limit and drift inwards across the limit due to aerodynamic gas drag. Their simulations show that all planetesimals that are drifting inside the stability limit are ejected. Especially for close binaries, the snow line is often exterior to the stability limit such that a considerable portion of their ejected planetesimals may be refractory. However, this mechanism is only efficient if a significant fraction of the solid disk mass drifts interior to the binary stability limit. For the aerodynamic forces to work efficiently the planetesimals have to be relatively small ($r <$ 1 km). Only then can they enter the dynamically unstable zone close to the binary, rather than piling up at the pressure maximum in the gaseous circumbinary disk. It has been proposed that 1I/`Oumuamua could be a fragment of a planet that formed in a binary system and was tidally disrupted after passing too close to one of the stars \citep{Cuk:2018}. However, tidal disruption is a rare event. Nevertheless, tight binary systems can eject an amount of rocky material comparable to the predominantly icy material thrown out by single- and wide binary-star systems. It is estimated that the mass of ISOs ejected from binary systems is at least the same as by planet scattering \citep{Jackson:2018} and the ISOs leave their parent binary system with $\approx$ 6.2 km/s \citep{Adams:2005}.

\subsubsection{Relative importance of the different ISO formation mechanisms}
\label{sec:ejection:relative}

All mechanisms mentioned in this section produce ISOs. The question is which of them is/are the dominant ISO production mechanism(s). All the processes discussed above produce of the order of 0.1 -- 30 $\mearth$ per star. However, given the strong dependence on the assumed mass of the planetesimal reservoir, these numbers fail to give a clear winner. Thus different criteria have to be employed. The various models make different predictions concerning the properties of the ISOs. As soon as a large enough sample of ISOs is known, these model predictions can be used to identify the dominant ISO production mechanism. Differences are mainly in terms of the composition and the velocity of the ISOs. 

Most mechanisms lead predominantly to icy ISOs that formed outside the snowline in their parent systems. However, 1I/`Oumuamua did not show the cometary coma morphology one would expect from an icy ISO. Therefore, many mechanisms have been tested concerning their ability to produce refractory ISOs. In general, scattering processes tend to produce more asteroidal ISOs than stellar flybys, drifting from the Oort cloud and the processes at the end of the main sequence. However, even scattering processes produce mainly icy ISOs.

There are significant differences in the ejection velocity of the various processes \citep{Pfalzner:2021}. The slowest ISOs are those that drift away from the Oort cloud or are shed during the stellar post-main-sequence phase ($\approx$ 0.1-0.2 km/s). Stellar flybys lead to ejection velocities in the range of 0.5-2 km/s; even higher velocities are achieved by planet scattering ($\approx$ 4-8 km/s), only exceeded by processes where two or more stars kick out the planetesimals together. 

Most processes discussed have a connection between velocity and composition. Due to their closeness to the star, the refractory planetesimals require a higher velocity to become unbound than the icy planetesimals (unless, of course the rocky planetesimals resided in the Oort cloud of the host star). As soon as a statistically significant sample of ISOs is discovered, the combined information of their observed velocities and composition might help constrain the dominant production process. However, the observed ISO velocity is only indirectly linked to the ejection velocity, and other effects such as dynamical heating for older ISOs may make such studies difficult. The complete ISO velocity distribution contains multiple components reflecting various ejection speeds and the parent system's different ages. In future, disentangling the different components will be one of the significant challenges.

\subsection{Resulting ISO population}
\label{sec:ejection:population}

Most of the above described planetesimal ejection mechanisms produce considerably more icy ISOs than rocky ones. The ejection of volatile-free asteroids is possible, but they are thought to be a relatively small fraction of planetesimals that were ejected. However, this impression might be partly caused by our heliocentric view. In the solar system giants planets all orbit beyond the “snowline”. Therefore, the majority of bodies within $<$2.5 au on unstable orbits end up colliding with the Sun
\citep{Gladman:1997,Minton:2010}.  In the solar system system there is indeed a much greater supply of icy planetesimals, however, this may not apply elsewhere \citep{Cuk:2018}. If we look at the known exoplanet systems, many of them have relatively massive planets much further in than in our solar system. The ejection of rocky planetesimals would be more efficient where there are close-in massive planets around high-mass stars. However, even then it appears unlikely that the rocky planetesimals could dominate the galactic population of scattered ISOs. The reason is that the large reservoir of volatile planetesimals beyond the snowline is always much larger than that of the rocky planetesimals in the inner systems. 

Naturally, the size distribution of ISOs is connected to the size distributions of the reservoir of planetesimals the ISOs originate from. Unfortunately, there is still a large uncertainty of the size distribution of planetesimals. Often a mass function of the form $N_{ISOs} \propto m^{-p}$ is assumed. This means that the number of ejected particles of a certain mass $N(m)$ is then given by
\begin{equation}
N(m) = \frac{2-p}{p-1}\frac{M_{ISO}}{m^{p-1}m_{up}^{2-p}}
\end{equation}
\citep{Adams:2005}, where $M_{ISO}$ is the total mass of ISOs and $m_{up}$ the largest mass possible. Sometimes $p$ = 5/3 is chosen \citep{Adams:2005}. \textcolor{black}{However, starting from a Dohnanyi distribution and from a distribution obtained by simulations of the streaming instability (p=1.6), \citet{Raymond:2018} argue that the underlying masses are off by 2-4 orders of magnitude.} Observational constraints of the small body population of the solar system can also provide some constraints on the initial size distribution of ISOs. Beyond Neptune, the small body population is thought be the least collisionally evolved \citep{Abedin:2022}. For these planetesimals a turnover in the size-number distribution is observed around a radius of  50 -- 70 km \citep{Bernstein:2004, Fraser:2014}, which likely is of primordial origin \citep{Vitense:2012}.  Therefore, one can expect that size distribution of the ISOs before expulsion was similar.

\begin{figure}[ht!]
\begin{center}
\includegraphics[width=8.5cm]{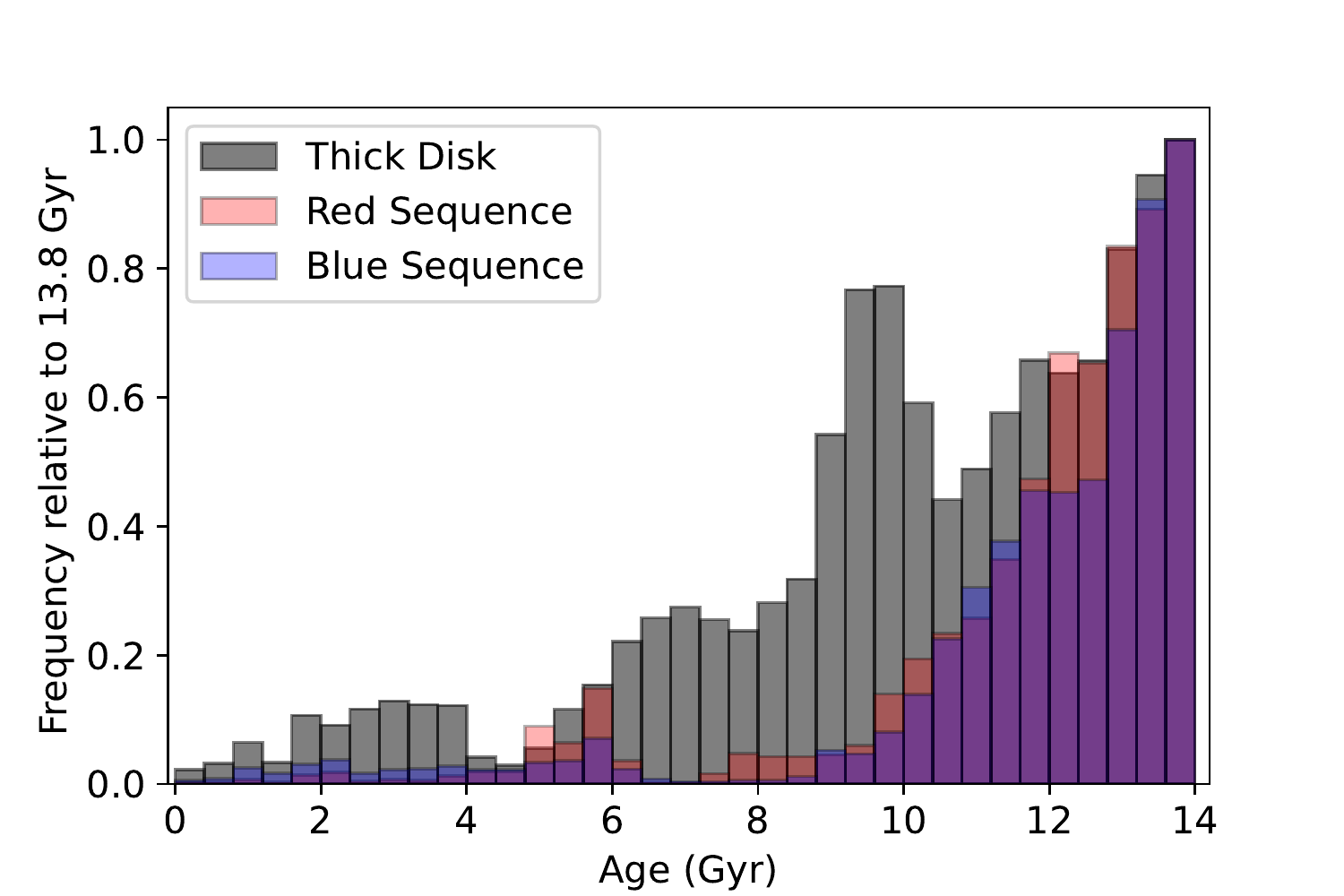}
\includegraphics[width=8.5cm]{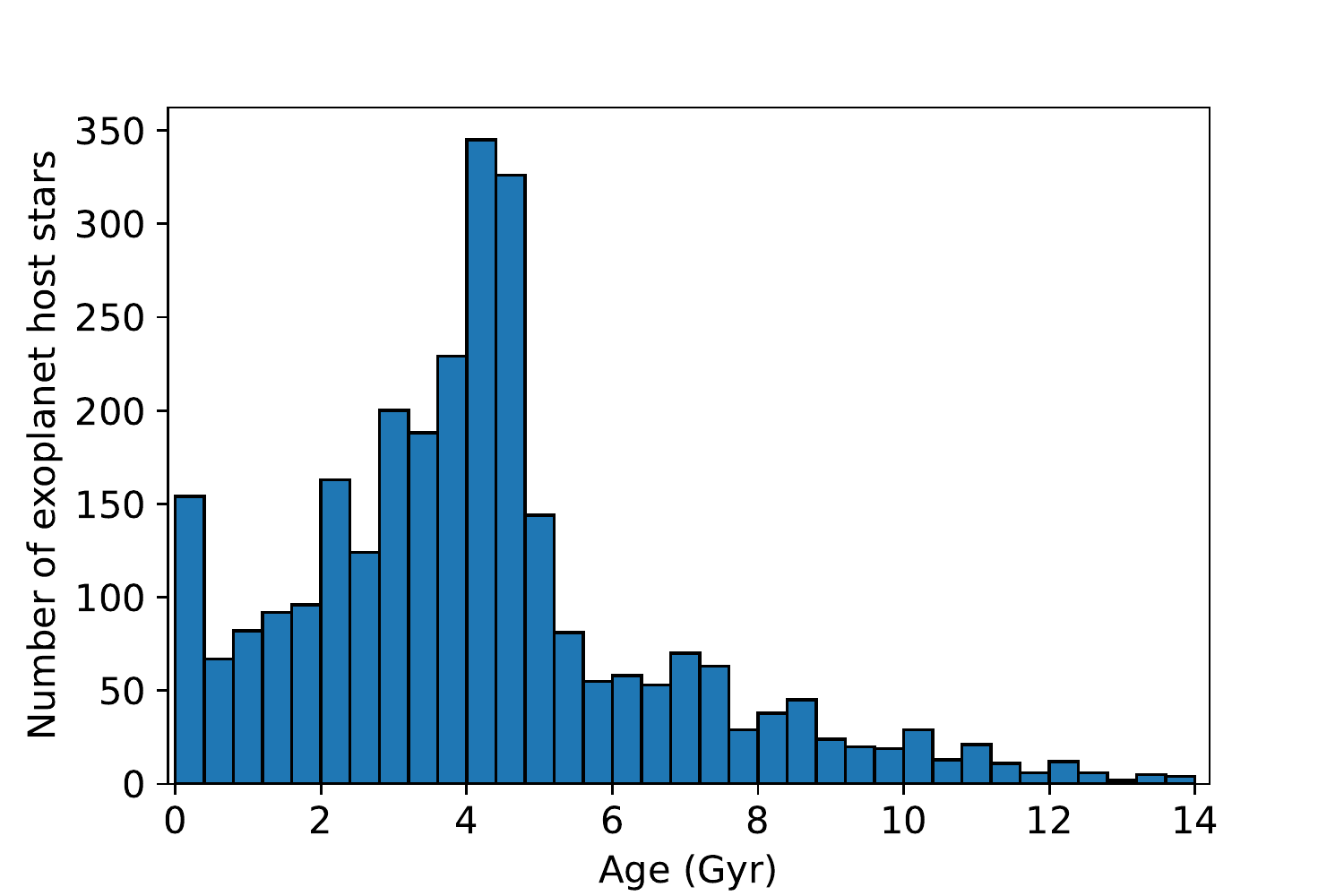}
\caption{Top: Stellar age distributions in the Milky Way, adapted from \cite{Gallart:2019}. Bottom: Age distribution of planet hosting stars as listed in the public database at {\it exoplanets.eu}. The observational bias towards surveying Sun-like stars is clearly evident. }
\label{fig:stellar:ages:milkyway}
\vspace{-0.25cm}
\end{center}
\end{figure}

\textcolor{black}{The size distribution of the planetesimals before expulsion sets the stage for the ISO size distribution. However, both distributions are not necessarily identical.} First, there might be selection effects concerning the planetesimals mass (and therefore size) in some of the ISO production mechanisms. For example, in three body scattering processes, ejection is more likely the lower the mass of the body. Second, the ejection processes themselves can alter ISOs, such that the properties of the ISOs differ from those of the planetesimals in the parent system. While ISOs that are relatively gently released from their parent system during the Oort cloud phase will remain unaltered, ISOs ejected in a more violent fashion might be disrupted during ejection \citep{Raymond:2020}. For example, planetesimals gravitationally scattered in very close encounters with a giant planet can be subjected to tidal disruption \citep[e.g.][]{Asphaug:1996}. In particular, scattering events that include a close passage to the parent star can lead to the decline or loss of cometary activity \citep[e.g.][]{Levison:1997}, and sublimation-driven activity may also flatten future ISOs’ shapes \citep{Seligman:2020, Zhao:2021}. Third, ISOs might be affected to different degrees by their journey through interstellar space (see \S\ref{sec:ISM}). 

By now, it should have become clear that planet formation and the production of ISOs are intrinsically linked processes. Ever since planets formed, a population of planetesimals were produced at the same time. A considerable portion of these planetesimals are released sooner or later from their parent star(s) and became ISOs. This means that the interstellar medium is steadily enriched with newly released ISOs \citep{Pfalzner:2019}.  Just like the Galaxy’s stellar metallicity increases over time so does the density of interstellar objects.

The ISO age distribution should be directly linked to the age distribution of the planets and stars. The stellar age distributions within the Milky Way are shown in Fig.~\ref{fig:stellar:ages:milkyway}. It demonstrates that star formation peaked at the early ages of the Galaxy. However, it is unclear whether the first stars produced ISOs at the same rate as today. It is known that planet formation is much more efficient around high-metallicity stars.  ISO formation is linked to planet formation. Suppose we take the age distribution of the planet host stars (Fig.~\ref{fig:stellar:ages:milkyway} bottom) as a proxy for the planetesimal production distribution, this differs considerably from the general stellar age distribution in the Milky Way.  The age distribution of planet host stars shows a clear maximum at approximately 4 Gyr. The age distribution of ISOs might shift even more to younger ages, as some portion of the ISOs may dissolve during their interstellar journey since the frequency of volatile loss and extinction is far higher for ejected planetesimals than for surviving ones \citep{Raymond:2020}.

\textcolor{black}{
\subsection{ISO timescales}
\label{sec:ISO_timescales}
A planetesimal that has been ejected from its host star and has become an ISO does not necessarily remain an ISO forever.  ISOs may be recaptured. Recapture is most likely if the planetesimal has been ejected from a host star which is a member of a star cluster.
This situation is most likely for young host stars because most stars are born in a star cluster environment \citep{Lada:2003}. However, some stars remain for hundreds of Myr in open clusters and for several Gyr in globular clusters.  In all these situations, recapture is a real option.}

\textcolor{black}{In such an environment, the ISO might be very quickly ($< 1$ Myr) recaptured by a different member of the same star cluster and lose its status of an ISO \citep{Hands:2019}. The likelihood of the ISO being recaptured rises with the cluster mass as the escape speed from the cluster increases with cluster mass.  Equally, ISOs escaping from stars in the cluster centre are more likely to be recaptured. For ISOs that are not in a star cluster environment, the likelihood of recapture is much lower. In any case, the velocity of the ejected ISO is the key parameter for being recaptured. The lower the ISO's velocity the higher the probability that it is recaptured. This correlation is valid in every capture environment, whether it is another star, a disc surrounding a young star or a molecular cloud \citep{Grishin:2019,Pfalzner:2021,Moro_Martin:2022}.}

\subsection{Physical processing in the ISM}
\label{sec:ISM}

ISOs are often viewed as travelling through interstellar space in calm cryogenic conditions. Therefore, ISOs are expected to be pristine samples of the planetesimals of other planets. As planetesimals are leftovers from the planet formation process, ISOs should therefore give us direct information about planet formation elsewhere. Generally, an ISO's journey through interstellar space might be not as uneventful as often imagined. The environmental effect on ISOs can be expected to be similar to that of Oort cloud comets. Just like the comets residing in the Oort cloud \citep{Stern:2003}, a variety of thermal, collisional, radiation, and ISM processes might affect ISOs during their long journey \citep{Stern:1990b}. The ISOs' surface is subject to these environmental influences, which can potentially lead to the erosion or even complete destruction of an ISO. How destructive these processes are depends on the composition of the ISO, with a rocky ISO being much less affected than, for example, a hydrogen-rich ISO.

During their journey, ISOs experience cosmic microwave radiation and radiation from the stars. The temperature increase by the cosmic microwave radiation on the surface of the ISO is probably negligible. In contrast, the effect of stellar radiation depends strongly on the ISO's individual journey. Any ISOs passing close to a star can be expected to be modified considerably by such an encounter. The prime effect is the reduction or complete loss of volatiles. Such an event is expected to be rare. However, parsec-range and closer encounters with highly luminous O and supergiant stars can heat ISO surfaces to temperatures capable of removing the most volatile ices, such as neon or oxygen \citep{Stern:2003}.  

ISOs are also subjected to the effects of cosmic rays and stellar energetic particles. Both have a broad spectrum of energies and interact with the ISO surface and subsurface. Cosmic rays are the primary source of space weathering for the comets in the Oort Cloud and, therefore, likely also for ISOs.  While low-energy particles interact only with the ISO surface, the most energetic ones deposit a significant amount of energy down to tens of meters. These processes can modify the isotopic ratios in cometary ices and create secondary compounds through radiolysis \citep{Gronoff:2020}. \textcolor{black}{The penetration depth of cosmic rays is relatively small in comparison to the ISOs of size $\geq 100$~m  that are likely to be detected by telescopes, see \S\ref{sec:searches}.} Even for the extreme case of pure hydrogen ISOs, it would be $<$ 10 m \citep{Hoang:2020}, for rocky material considerably shallower. Thus cosmic rays immediate effect on ISOs is small, but over time it can erode an ISO gradually from the surface or affect the composition of the upper layers (see \S\ref{sec:origins}). 

The interstellar medium contains large amounts of gas and dust. The effect of this gas and dust on ISOs depends on the gas and dust densities and the relative velocity of the individual particles. Gas and dust densities are relatively low in the interstellar medium itself ($n_H \approx$ 10 cm$^{-3}$) and are considerably higher in molecular clouds ($n_H \approx$ 10$^4$ cm$^{-3}$) and even higher when passing through protoplanetary disks of newly forming stars ($n_H \approx$ 10$^{5-7}$ cm$^{-3}$).  

Collisions of ISOs with the ambient gas at high speeds can heat the frontal area, possibly resulting in transient evaporation. Such a situation occurs when the ISO passes through a molecular cloud. Each particle directly impacting on the surface of an ISO delivers an energy $E_p=m_p v_p^2/2$. Therefore, ISOs are most impacted by high-velocity particles.  At least for Oort cloud comets, simulations show that dust particles could significantly erode the cometary surface in the range of several meters \citep{Stern:1990b, Mumma:1993}, preferentially removing the sub-micron grains. However, erosion by ISM grains is a complex process depending on the ISM grains composition and structure.

While moving through a relatively massive molecular cloud forming a large star cluster, there is a non-negligible probability that the most massive stars explode as a supernova. Therefore, an ISO might be subjected to the explosion products from such a supernova. Despite supernovae explosions being brief ($\approx$ 0.1 yr), they are extremely luminous $L= 10^9 L_{sun}$ and therefore can heat ISOs from considerable distances. Their intense but shorter thermal pulses could propagate 0.1--2 m into ISO surfaces \citep{Stern:2003}.

These processes complicate the interpretation of ISO observations; the surface layer of ISOs might be modified over time. In summary, all these processes erode the surface of ISOs to different degrees. However, as the distinctive properties of the ISOs in terms of their size and composition are not well constrained, it remains unclear whether their surface is affected or whether they decrease in size with age and how many small ISOs become destroyed.  

\begin{figure*}[hb!]
\begin{center}

\includegraphics[width=17cm]{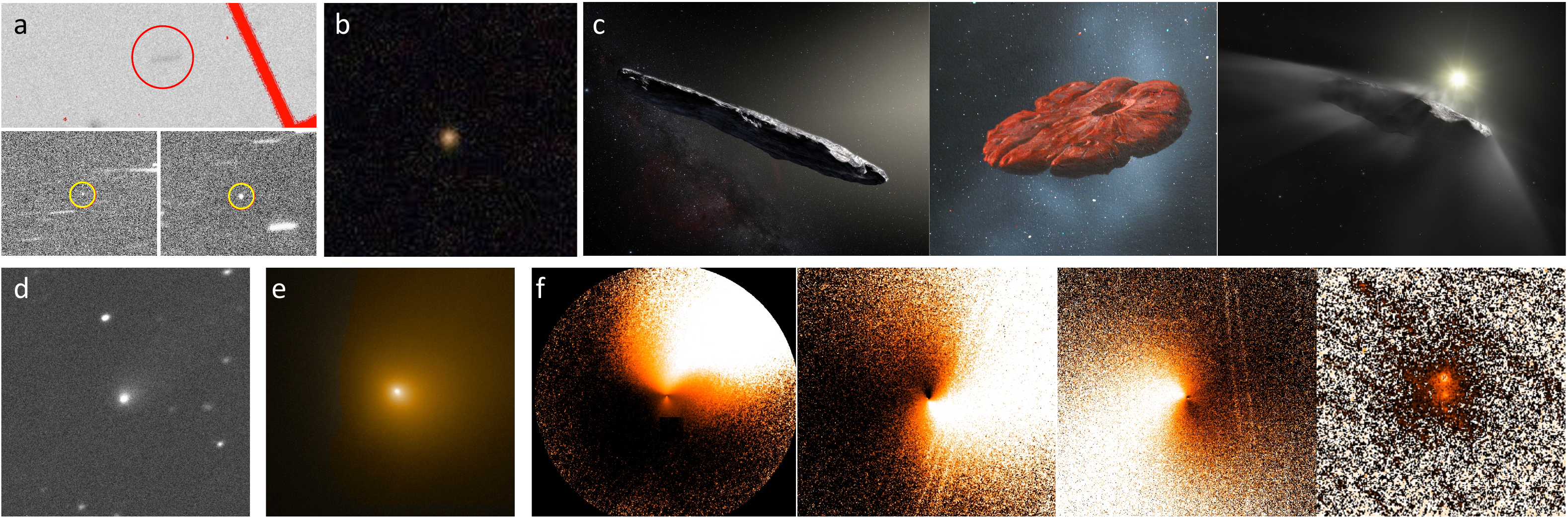}
\caption{Images of the two ISOs passing through the inner solar system show the dramatic difference in their characteristics. Top row: 1I/`Oumuamua: (a) Pan-STARRS discovery image (top) taken on 2017 Oct. 19, with follow up images from the CFHT near min/max brightness on Oct. 27; (b) Gemini 8-m color composite image made from 192 images (1.6 hrs) showing no hint of a dust coma (Gemini Observatory/AURA/NSF); (c) artist's depiction of two possible nucleus shapes based on the large light curve range (Credit: ESO, M. Kornmesser; William Hartmann) and artist's view of the ISO after the discovery of non-gravitational acceleration (ESA/Hubble, NASA, ESO, M. Kornmesser). Bottom row: 2I/Borisov: (d) CFHT image taken 10 days post-discovery (2019 Sep. 10); (e) HST image on 2019 Oct. 12 (NASA/ESA); (f) radially normalized ratio HST images median averaged over an orbit from 2019 Oct. 12, Dec 24. (00:31 UT), and Dec 24. (02:21 UT) showing outgassing jets, and 2020 Jul. 6 showing the split nucleus (NASA/ESA), processed by H. Boehnhardt.}
\label{fig:1I-2I}
\end{center}
\end{figure*}
\section{\textbf{ISO SEARCHES AND DISCOVERY}}
\label{sec:searches}

\subsection{Dynamically recognising ISOs}

\textcolor{black}{Upon discovery, ISOs will appear as either comets or asteroids depending on their level of activity. Hence the single most important discriminant is showing the orbit is hyperbolic, with $e>1$ and $a<0$ at the 3-$\sigma$ level or higher. Dynamically, this is equivalent of saying that their velocity at an infinite distance $v_\infty>0$ (also known as the hyperbolic excess velocity).  This requires accurate astrometric measurements of the object over a period of time, and hence sensitive wide-field surveys that allow early detection. For ISOs close by the Earth it may take only a few days of observations before the orbital arc is large enough to show this unambiguously; more distant ISOs may take weeks (see \S\ref{sec:oumuamua}).}

\textcolor{black}{It should be noted that many comets are discovered each year on apparently hyperbolic orbits. These are in reality dynamically new comets from the Oort cloud, but are listed as hyperbolic due to a number of reasons. First, an orbit that is near-parabolic in the solar system barycentric reference frame may appear to be hyperbolic in the heliocentric reference frame, in which most orbital elements are published. For example, the dynamically new comet C/2021 A9 (PANSTARRS) has an osculating heliocentric orbital eccentricity of $e=1.004$. Transforming to the correct barycentric frame gives $e=1.0004$, much closer to a parabolic orbit. Secondly, an originally weakly bound object may become unbound due to gravitational perturbations or non-gravitational forces due to outgassing, see \cite{Krolikowska:2010}.}

\textcolor{black}{ To illustrate this, using data from the JPL Small Bodies Database 
( https://ssd.jpl.nasa.gov/tools/sbdb\_query.html ), 17 Long-Period Comets (LPCs) were discovered in 2020--2021 with osculating orbital eccentricity $e\geq 1.0$. Yet all eccentricities were so close to unity that either these comets have $e\leq 1.0$ within the orbital uncertainties and/or within a barycentric reference frame, or they would have possessed an originally parabolic/elliptical orbit before entering the our planetary system. Understanding these factors is crucial for confirming the status of suspected ISOs with $e\simeq 1$.}

\subsection{ISO Studies pre-1I/`Oumuamua}

There were many papers that made predictions about the space density of ISOs based on the non-detection in surveys. Some very early work  was reported by  \cite{Safranov:1972, McCrea:1975, Sekanina:1976, Duncan:1987, Valtonen:1990, Zheng:1990}. The advent of wide-field CCD-based sky surveys in the 1990s allowed quantitative upper limits to be assessed. \citet{Francis:2005} derived a 95\% upper confidence limit of $4.5\times 10^{-4}$ au$^{-3}$ using the LINEAR survey. Just before the discovery of 1I/`Oumuamua, \cite{Engelhardt:2017} numerically integrated ISOs and linked them to the non-detection by the combined major sky surveys operational at that time; the Catalina Sky Survey, the Mt. Lemmon Survey and the Pan-STARRS project. Although they assumed isotropic approach trajectories which disregarded the local standard of rest distribution of stellar velocities, they presciently derived upper limits for both inactive and active (cometary) ISOs. They reported 90\% confidence upper limits for 1 km diameter ISOs of $\leq 2.4\times10^{-2}$ au$^{-3}$ and $\leq 1.4\times10^{-4}$ au$^{-3}$ respectively.

The important thing to get across here is why was the first ISO discovered now?  To discover these fast moving objects you need to survey the whole sky quickly it to very faint limiting magnitudes, and we have only had surveys capable of doing this in the last couple of decades. \cite{Heinze:2021} shows that current near Earth object (NEO) surveys are poorer than expected at finding fast moving objects. However, it is expected that future surveys will do much better (see \S\ref{sec:VCRO}). 

\section{\textbf{OBSERVED CHARACTERISTICS OF ISOs}}
\label{sec:oumuamua}

\subsection{1I/`Oumuamua}
Some of the most important questions about the first ISO include: (1) Where did it come from? and (2) What is it made of? \textcolor{black}{As discussed below,} we still don't know the answers to these questions.  1I/`Oumuamua was easily observable from the ground for a little over a week, but as it moved away from the Earth it faded quickly, and large telescopes were able to observe for only about 1 month. The last Hubble Space Telescope observations were made in Jan. 2018. It is remarkable that we know as much about this object as we do, because all of the large telescope time had to be secured through Director's requests. In total approximately 100 hrs on 2.5-10-m ground-based telescopes were devoted to characterizing this exceptional object. \textcolor{black}{To date, over 200 refereed papers have been written on both interstellar objects, and nearly 450 papers including non-referred material.  It is not practical to cite everything, so key papers have been highlighted, and we have included some reviews of the field.}

\subsubsection{Discovery of 1I/`Oumuamua}

On 2017 Oct. 19 the Pan-STARRS survey found an object, designated P10Ee5V, moving quickly with respect to the stars (see Fig.~\ref{fig:1I-2I}a) during the normal near earth object survey. Rob Weryk then found pre-discovery images in Pan-STARRS data from Oct. 18. Follow up data obtained by the 1m ESA ground station on Oct. 20 was rejected by the Minor Planet Center because the data implied a large eccentricity, and P10Ee5V was classified as an Earth-orbit crossing asteroid. Data from Oct. 20 obtained by the Catalina Sky Survey suggested that the object should be classified as a short period comet. However, observations obtained from CFHT on Oct. 22 showed that the orbit was hyperbolic with an eccentricity of 1.188. On Oct. 24 the object was designated C/2017 U1 (MPEC 2017-U181). This was corrected on Oct. 26 after deep images from the CFHT from Oct. 22 showed no coma. The object was named A/2017 U2 (MPEC 2017-U183).

Within a week of discovery a request was made to the Hawai'ian cultural group that advised Maunakea observatories management group to propose a name for the new object. They proposed the name `Oumuamua, meaning ``a messenger from afar arriving first'' or a scout or messenger sent from our distant past to reach out to us or build connections with us. In an extraordinarily fast effort, the IAU approved this on 2017 Nov. 6, and the new name became 1I/`Oumuamua. This became the foundation of a project called A Hua He Inoa which blends traditional indigenous practices into the official naming of astronomical discoveries \citep{Kimura:2019, Witze:2019}.

\subsubsection{Nuclear Characteristics}

One of the most straight forward measurements to make is the brightness, and from this get an estimate of the object's radius for an assumed albedo:
\begin{equation}
\label{eq:nucleus}
pr_N^{2} = 2.235 \times 10^{22} r_h^{2}
\Delta^{2} 10^{[0.4(m_{\odot}-m)]} 10^{0.4(\beta\alpha)}
\end{equation}

\noindent
where $r_h$ and $\Delta$ are the helio- and geocentric distances [au] and m$_{\odot}$ and m are the apparent magnitudes of the sun and comet \citep{Russell:1916}. It was apparent from the earliest imaging observations of 1I/`Oumuamua that it had a very large rotational light curve range so various estimates of the size were reported, even for the same assumed albedo. Combining several nights worth of observations from several observatories yields $H_V$ = 22.4 which gives an average radius of 0.11 km assuming a typical cometary albedo of 0.04 \citep{Meech:2017}. {\it Spitzer} observations of 1I/`Oumuamua, in principle could have provided measurement of both the nucleus size and albedo, however because of strict solar avoidance angles these observations could not be made until late 2017 Nov. and only upper limits on the flux were obtained \citep{Trilling:2018}. Thermal models were used to estimate the corresponding radius and albedo, with a preference for an albedo of 0.1 and a radius consistent with the previous estimate.

\begin{figure*}[ht!]
\begin{center}
\includegraphics[width=15.5cm]{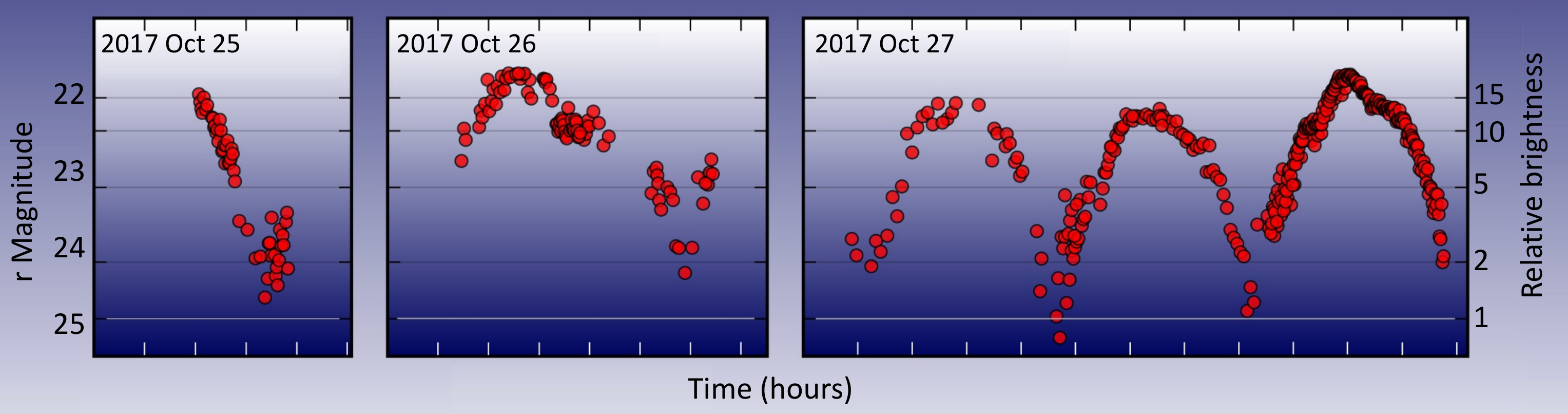}
\caption{Rotational light curve of 1I/`Oumuamua compiled from data taken from various observatories in 2017 October.}
\label{fig:lightcurve}
\end{center}
\end{figure*}

The rotational light curve shown in Fig.~\ref{fig:lightcurve} has a brightness range of $\Delta$m $\sim$ 3 mag, implying an axis ratio for an oblate sphere of \mbox{$\frac{a}{b}$ = 10$^{0.4(\Delta m)} = 15$.} However, this does not take into account the phase angle (which can make the object look more elongated than it is  (see Fig.~\ref{fig:1I-2I}c)). It also does not take into account that the rotation pole position is completely unknown; if the pole was more closely aligned to the ecliptic, then the object could appear even more elongated that the light curve suggests. The consensus is that the ratio is likely a/b $\gtrsim$ 6 \citep{ISSIteam:2019, Mashchenko:2019}. Solar system objects typically do not have axis ratios this large, and the cause of the unusual shape of 1I/`Oumuamua remains one of the enduring mysteries (see \S\ref{sec:origins}).
 
Attempts to find a rotation period from the data produced a variety of periods, near 8 hrs, all differing depending on the length of the data set. The most comprehensive model concluded that 1I/`Oumuamua is in a complex rotation state, rotating around its shortest axis with a period of 8.67$\pm$0.34 hours with a period of rotation around the long axis of 54.48 hours \citep{Belton:2018}. The damping timescale for a body this small is long enough that an excited rotation state can be preserved from the time of ejection from the host star. The shape of 1I/`Oumuamua as interpreted from the rotational light curve can change significantly, depending on the rotation state (see Fig.~\ref{fig:1I-2I}c).

\citet{McNeill:2018} used the shape and rotation of 1I/`Oumuamua to place some constraints on the strength and density of the ISO, finding it was more likely to have a density typical of asteroids.

\subsubsection{Constraints on composition}

Several groups obtained spectra for 1I/`Oumuamua, both from the ground and from {\it Spitzer}, but there was no evidence of outgassing (see \citet{ISSIteam:2019} for a summary). Many groups used either spectra or filter photometry to estimate the spectral reflectivity of the surface. The spectral reflectivity, $S$, was found to increase with wavelength and ranged between $S$ = (7-23)$\pm$3\%/1000 \AA~\citep{Jewitt:2022}. This measurement was challenging because the rotational signature had to be removed. Regardless of the value, the surface of 1I/`Oumuamua was red, typical of organic-rich comet surfaces, but also consistent with iron-rich minerals and space weathered surfaces.

Deep stacks of images from many nights of data from ground based and space-based optical images showed no dust at all to a limit of there being $\lesssim$ 10$^{-3}$ kg/s of micron-sized grains being produced \citep{Meech:2017,Jewitt:2017}.  In contrast, the dust production upper limit from {\it Spitzer} observations for 10 $\mu$m grains was $<$ 9 kg/s \citep{Trilling:2018}.

\subsubsection{Trajectory and potential origins}

One of the primary goals for the HST time awarded for 1I/`Oumuamua was to extend the astrometric arc length to be able to determine its orbit with sufficient precision to trace its path backwards and determine its home star system. \textcolor{black}{The chances of being able to trace the trajectory backwards to find its parent star were low because gravitational scattering from stellar encounters would limit the search to the past few tens of millions of years \citep{Zhang:2018}.} 

As 1I/`Oumuamua faded, astrometry was obtained with the VLT in Chile and with HST, with the last observations from HST being obtained on 2018 Jan. 2 when the ISO was fainter than V $\sim$ 27 at 2.9 au.  An analysis of the combined ground and HST observations showed that the orbit could not be well fit with a gravity only solution.  By adding a radial non-gravitational acceleration term directed radially away from the Sun with a $r_h^{-1}$ dependence, the orbit was well fit \citep{Micheli:2018}. \textcolor{black}{The authors ruled out a number of other more  hypotheses, e.g. Yarkovsky effect, frictional drag forces, impulsive velocity changes, photocenter offset from a binary or non-uniform albedo, interaction with the solar wind or radiation pressure.  Most were rejected because the acceleration produced by these mechanisms was either too small, in the wrong direction, wouldn't match the continuous nature, or implied a non-physical bulk density for \Ou.} By assuming that the same non-gravitational acceleration was operating on the pre-perihelion trajectory, an attempt was made to find the originating star system for 1I/`Oumuamua. \textcolor{black}{\citet{Dybczynski:2018} attempted a search for the parent star taking advantage of the {\it Gaia} satellite data release 1 (DR1) of stellar positions, but did not find an obvious parent star for \Ou.}

The {\it Gaia} DR2 data release provided the necessary astrometric position and {\it velocity} information that was needed to be able to try to trace the path of 1I/`Oumuamua back to its home star using its Keplerian orbit \citep{Bailer-Jones:2018}. Plausible home stars would be those which had been within 0.5 pc of 1I/`Oumuamua's trajectory (the approximate size of the Sun's Oort cloud), and which had low encounter (ejection) velocities $<$ 10 km/s (see \S\ref{sec:ejection:individual}). An initial selection of stars which passed within 10 pc of the trajectory was selected and the candidates and the path of 1I/`Oumuamua were integrated through a smooth Galactic potential. This resulted in four potential ``home'' systems with encounter distances between 0.6-1.6 pc and encounter velocities between 10.7-24.7 km/s, all with ejection times less than 6.3 Myr ago. Unfortunately, all of these systems had ejection speeds much higher than expected for ISOs.  

\subsubsection{Cause of the non-gravitational acceleration}

In order to explain the non-gravitational acceleration of 1I/`Oumuamua in the absence of any apparent activity (dust coma or gas), \citet{Micheli:2018} estimated the mass loss needed to accelerate the nucleus to the observed value for a range of comet and asteroid densities. They found mass loss requirements between 0.7-140 kg/s, but adopted 10 kg/s as the best estimate.  A thermal model matching the acceleration would require sublimation of water plus a more volatile species (such as CO$_2$ or CO).  With a high CO/H$_2$O ratio the model could produce enough mass loss to within a factor of 2-3 of what was needed to accelerate 1I/`Oumuamua. \citet{Trilling:2018} found 3$\sigma$ upper limits for the dust production of 9 kg/s (for 10 $\mu$m grains), $Q$(CO$_2$) = 9 $\times$ 10$^{22}$ mol s$^{-1}$ and CO 1.1 $\times$ 10$^{24}$ mol s$^{-1}$ (after correction). None of the optical spectra were sensitive enough to detect gas with such a low CO or H$_2$O production rate. Having consistency within a factor of a few between the model and observations suggests that volatile outgassing is a plausible scenario for the acceleration of 1I/`Oumuamua.  \citet{Stern:1990b} postulated that any comet passing through a molecular cloud would have the small grains eroded from the surface, and this could explain the lack of dust in the presence of outgassing.

\textcolor{black}{Other more exotic volatile sublimation scenarios to explain the non-gravitational acceleration are discussed in \S\ref{sec:origins}.}
Finally, \citet{Seligman:2021} investigated the spin dynamics of 1I/`Oumuamua with outgassing consistent with the non-gravitational acceleration and found that this need not cause a spin up of the nucleus. They were able to reproduce the observed light curve with a model of CO outgassing but not with water-ice jets.

\subsection{2I/Borisov}
\label{sec:borisov}

\subsubsection{Discovery of 2I/Borisov}

The second ISO was discovered on 2019 Aug. 30 by Gennadiy Borisov, an engineer at the Crimean Astronomical Station and amateur astronomer, using a 0.65m f/1.5 astrograph \citep{Borisov:2021}. Borisov reported it as a potential new comet with a compact 7 arcsec diameter coma, and detected a short 15 arcsec tail two days later. It was placed  on the Minor Planet Center Potential Comet Confirmation Page as object gb00234, and by 2019 Sep. 11 sufficient additional astrometry had been reported to give a cometary designation of C/2019 Q4 (Borisov) and an initial orbit with an eccentricity $e=3.08$ (MPEC 2019-R106). On 2019 Sep. 24 the permanent designation of 2I/Borisov was announced (MPEC 2019-S72). As further observations were reported, by mid-October the derived orbit had stabilized with $e=3.354$, $q=2.006$ au and $i=44.05^\circ$, giving discovery circumstances of $r_h=2.99$ au and $\Delta=3.72$ au. This allowed \cite{Ye:2020} to identify precovery images in archival survey data dating back to December 2018 when it was at a heliocentric distance of 7.8 au.

\subsubsection{Trajectory and potential origins}

The significantly longer observation arc for 2I/Borisov allowed the measurement of standard cometary  non-gravitational forces, implying that searches for progenitor systems might suffer less uncertainty than for 1I/`Oumuamua. However, \citet{Hallatt:2020} did not identify any systems with the golden combination of small encounter distance and low encounter velocity. \citet{Bailer-Jones:2020} investigated a range of non-gravitational force models and found that 2I/Borisov was $\sim0.053-0.091$ pc from the M0V star Ross 573 910 kyr ago. They also demonstrated that such a close encounter is unlikely over this short time period. However the predicted encounter velocity was $22.6$ km/s, significantly higher than expected for ejection of ISOs, even assuming giant planet interactions \citep{Bailer-Jones:2018}. 

Hence like 1I/`Oumuamua, the home system of 2I/Borisov remains unidentified. It is likely that unless the ejection was from a nearby star very recently, that tracing the path to the home system is not possible. As soon as an ISO passes through a molecular cloud, back-tracking no longer possible.

\subsubsection{Nuclear Characteristics}

Due to its activity, there were no direct detections of the nucleus of 2I/Borisov. In their archival analysis, \citet{Ye:2020} did not detect 2I/Borisov in deep co-added images from November 2018. This gave a strong upper limit to the nucleus radius of $r_N \leq 7$ km assuming a geometric albedo of $0.04$. Rigorous imaging constraints came from HST observations by \cite{Jewitt:2020a}, who measured a strong upper limit of $H_V\geq16.60\pm0.04$ assuming a nominal cometary phase function of $0.04$ magnitudes/degree. Fitting the coma surface brightness distribution gave showed no significant excess from a central nuclear psf, giving an improved upper limit to the nucleus radius of $r\leq 0.5$ km, or $H_V\geq19.0$. \cite{Hui:2020} calculated the non-gravitational parameters for a range of heliocentric distance laws, and combined them with outgassing measurements (see below) and an assumed nuclear density of $\rho=0.5$ gm cm$^{-3}$ to derive $r\leq 0.4$ km. \cite{Xing:2020} used measured water production rates to derive a minimum radius of $r\geq 0.4$ km assuming the entire sunward surface of the nucleus was active.

Several authors used apparent jet structures visible from HST imaging of the inner coma to attempt to constrain the spin axis. \cite{Kim:2020} found it difficult to explain the overall dust coma anisotropy in terms of a single isolated emission source, and preferred a model of general mass loss from the afternoon nuclear surface, giving a pole orientation of $(\alpha, \delta)=(205^\circ, +52^\circ)$. From the same data, \cite{Manzini:2020} interpreted the appearance of two jet/fan structures as implying localized emission near the rotation equator, and derived a spin pole direction of $(260^\circ, -35^\circ)$. Combining the \citet{Jewitt:2020a} data with additional HST imaging, \cite{BolinLisse:2020} interpreted the positional evolution of the jets as implying a near-polar source with a direction $(322^\circ, +37^\circ)$. Although all three studies reported uncertainties of at least $\pm10^\circ$, the disparity between these results led \citet{BolinLisse:2020} to conclude that it was not possible to sufficiently constrain the pole orientation with the extant data. The latter study also reported a tentative detection of a 5.3~hr periodicity in the light curve, but the low amplitude of $\sim0.05$ magnitudes implies a low significance. The lack of precise determinations of these nuclear properties are all unremarkable when compared to remote studies of solar system comets (see {\it Knight et al.}, in this volume). 

The similarity of 2I/Borisov to solar system comets extended to the detection by Drahus {\it et al.} (ATEL 13549) of two outbursts on 2020 Mar. 4 and 8, brightening the comet by $\sim 0.7$ magnitudes. A sequence of HST imaging from 23 March to 20 April to investigate possible nuclear fragmentation was reported by \cite{Jewitt:2020b}. These images showed a discrete nuclear secondary on 30 March, 26 days after the first outburst, that was not visible in images 4 days later. They posited this transient nature by the secondary consisting of one or more boulders ejected from the nucleus during the outburst, disrupting weeks later near the time of their observations due to rotational spin-up.

\input{tab-1I-2I}

The measured properties of 1I/`Oumuamua and 2I/Borisov are summarized in Table~\ref{tab:1I-2I}.

\subsubsection{Composition and evolution}

The first characterizations of the coma of 2I/Borisov were obtained even before it was given a provisional designation, with early optical photometry and spectra of the coma giving a red reflectance spectrum typical of normal comets (\citealt{JewittLuu:2019, deLeon:2019,Guzik:2020}) and similar to 1I/`Oumuamua \citep{Fitzsimmons:2018}. Gas emission via the bright CN(0-0) emission band at 388~nm was first detected on 2019 Sep. 20-24 by \cite{Fitzsimmons:2019} and \cite{deLeon:2019}. Spectroscopy and narrow-band photometry over the following month was reported by \cite{Opitom:2019}.  As with the earlier spectroscopy, they did not detect the C$_2$(0-0) emission band at 517~nm and concluded the comet was similar to carbon-chain depleted solar system comets. However weak detection of C$_2$(0-0) was reported by several observers from November ({\it i.e.} \citealt{Lin:2020}), and by perihelion on 8 December 2019 the C$_2$/CN ratio was formally consistent with both depleted and normal carbon-chain cometary abundances (see Fig.~4 of \citealt{Aravind:2021}). These spectra and those presented by \cite{Opitom:2021} also showed other normal cometary molecular emission features due to species such as NH, NH$_2$, and CH. Hence the optical species showed a significant evolution during this time as the heliocentric distance decreased from 2.7~au to 2.0~au

The first constraint on the sublimation rates of primary ice species came from detection of the [OI] 6300 \AA\ line by \cite{McKay:2020}. Assuming this came solely from sublimation and subsequent dissociation of water, they derived a production rate of $Q(H_2O)=(6.3\pm1.5)\times10^{26}$ mol s$^{-1}$ on 29 October at $r_h=2.38$ au. \cite{Xing:2020} used the UVOT telescope onboard the NASA Swift satellite to observe the OH (0-0) and (1-1) emission between $280-320$~nm, again resulting from dissociation of water molecules. The derived water production rates showed a peak of $Q(H_2O)=(1.1\pm0.1)\times 10^{27}$ mol s$^{-1}$ near perihelion, but with higher production rates pre-perihelion than post-perihelion.

Given the significant activity at $r_h \geq 3$ au, it was clear that other volatile species such as CO or CO$_2$ should be present in 2I. \cite{Bodewits:2020} used HST to measure the CO fluorescence bands at 140--170 nm and corresponding production rates during and after perihelion. \cite{Cordiner:2020} used ALMA to measure CO ($J=3-2$) emission along with HCN ($J=4-3$). Both these teams showed that the CO abundance was exceptionally high in 2I/Borisov, being on par with the H$_2$O abundance at $Q(CO)/Q(H_2O) \simeq 0.35-1.55$. Only rare CO-rich comets such as C/2016 R2 had previously exhibited such CO-rich compositions (see \citet{McKay:2019} and the chapter by Biver {\it et al.} in this volume). 

\cite{Bodewits:2020} found that the CO production rate stayed relatively constant around perihelion, while that of H$_2$O fell soon after, implying a significant change in the near-surface abundance ratios of these ices. This may have been due to seasonal effects of nucleus insolation, coupled with  a non-heterogenous composition or erosion of the surface, similar to that seen by ESA Rosetta in some species at comet 67P/Churyumov-Gerasimenko \citep{Lauter:2020}, although for that comet the CO/H$_2$O ratio was relatively constant within 2.5~au. Similarly, as mentioned above, the abundance ratios of the primary optical species C$_2$/CN also changed significantly as the comet moved towards perihelion. Finally, shortly after the discovery of significant neutral metal line emission in optical spectra of non-sungrazing comets \citep{Hutsemekers:2021}, these were also found in spectra of 2I/Borisov \citep{GuzikDrahus:2021, Opitom:2021}. The abundance ratio of $\log (Q(NiI)/Q(FeI)=0.21\pm0.1$ was within the range measured for solar system comets, as was $Q(NiI+FeI)/Q(CO)$.

Taken together, although several authors concluded that the optically active species were consistent with normal solar system comet abundances near perihelion, these studies point to a significantly evolving coma composition as 2I/Borisov passed the Sun. It is unclear to what degree this evolution was caused by its interstellar nature, although the CO-rich composition is clearly unusual.

\subsection{ISO number densities and size distribution}

With the discovery of 1I/`Oumuamua and 2I/Borisov, astronomers could finally estimate the true space density of ISOs in the Solar neighborhood, albeit with large uncertainties given $N=2$. A complication also arises given their different apparent nature during their encounters, as to whether they may come from different populations of ISOs. Inert ISOs like 1I/`Oumuamua appear to have a local space density of  $\sim 0.1$ to $0.3$~au$^{-3}$  at $H=22$ \citep{Meech:2017,Do:2018}. This approximately agrees with pre-discovery upper limits for inert ISOs \citep{Engelhardt:2017} when scaled to this absolute magnitude. \textcolor{black}{Some authors reported a possible tension between the inferred mass density contained in ISOs and the galactic stellar density, with 1I/`Oumuamua implying an exceptional amount of mass-loss from planetary systems {\it e.g.} \cite{Rafikov:2018}.} However the \cite{ISSIteam:2019} demonstrated that the space density of $\sim 10^{15}$~pc$^{-3} \equiv 0.1 $~au$^{-3}$ was compatible with possible size distributions. 

The space density of ISOs exhibiting normal cometary activity like 2I/Borisov is less well constrained, even though most earlier upper limits {\em assumed} cometary activity. \cite{Engelhardt:2017} found a pre--2I/Borisov 90\% upper limit of $\leq 1.4\times 10^{-4}$~au$^{-3}$ for a nuclear absolute magnitude $H\leq 19$, assuming the entire sunward surface was active. \cite{Eubanks:2021} used the statistics of LPC discoveries, together with assuming that ISOs share the same velocity distribution as stars within 100~pc, to derive a lower density of $7\times 10^{-5}$~au$^{-3}$. This is only a factor 2  less than \cite{Engelhardt:2017}, and is plausible given the extra survey time since that study.

\begin{figure}[ht!]
\begin{center}
\includegraphics[width=8.3cm]{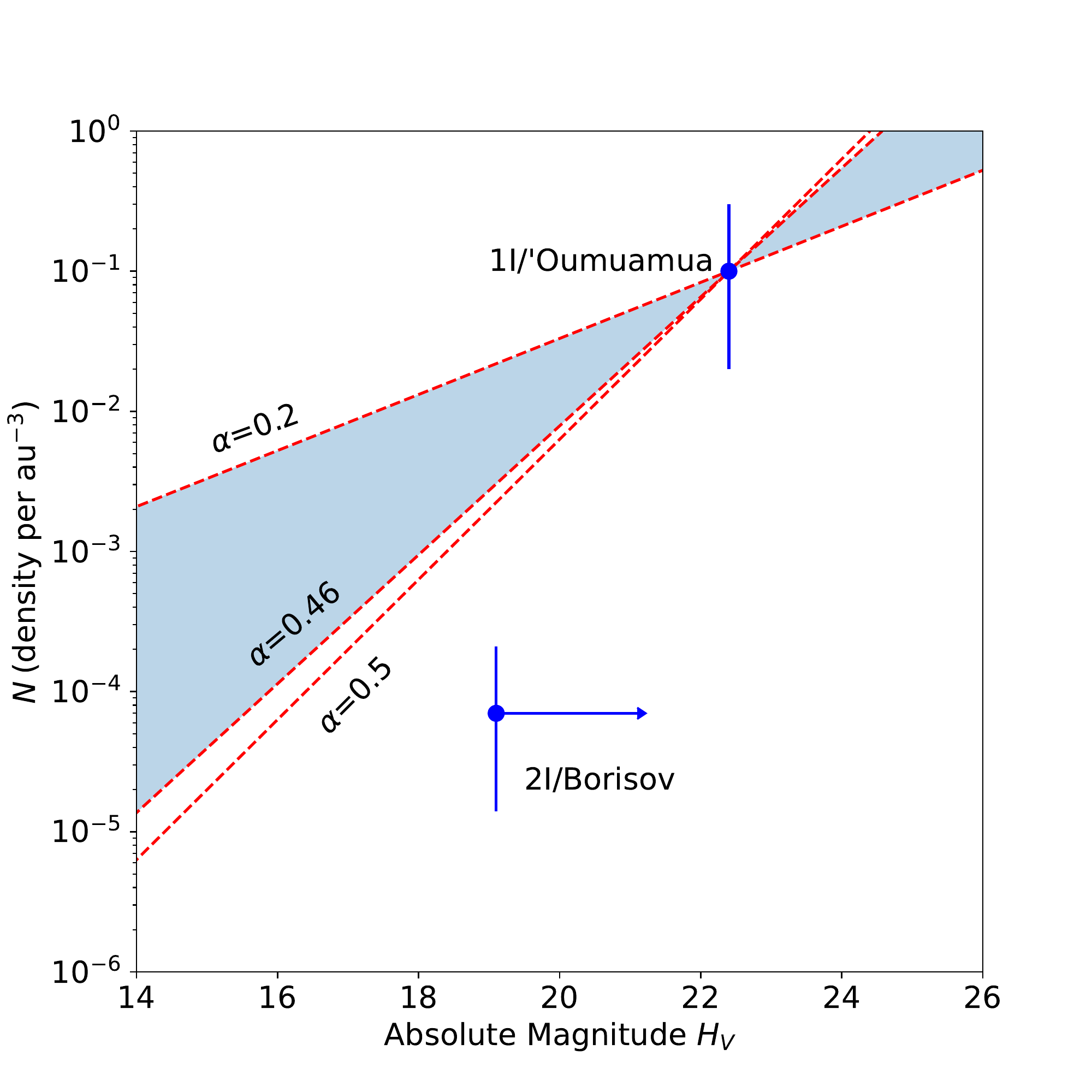}
\caption{Number densities for 1I/`Oumuamua and 2I/Borisov as a function of their absolute magnitudes. Also shown are differential brightness power-laws with exponents $\alpha=0.5$ for collisional equilibrium, and the range $\alpha=0.46$ to $0.2$ measured for solar system comets. All are anchored to the absolute magnitude of 1I/`Oumuamua.}
\label{fig:iso-sizes}
\end{center}
\end{figure}

If 1I/`Oumuamua and 2I/Borisov come from the same ISO population, then Fig.~\ref{fig:iso-sizes} provides constraints on the luminosity distribution. We use absolute magnitudes to avoid any assumptions on albedo, although it is important to remember these have been derived using assumed phase laws. If they share the same albedo then this also constrains the size distribution. For the interstellar space density from 1I/`Oumuamua we take the space density and confidence limits from \cite{Levine:2021}, and for 2I/Borisov we take the space density of \cite{Eubanks:2021} together with Poisson uncertainty ranges calculated for a single detection at both magnitudes.

We assume a single power-law relationship for the number density against size, as with only 2 points a more complex relationship is unwarranted. An obvious size distribution to test is the theoretical  differential size distribution for a population in collisional equilibrium, where the number of objects at a radius $R$ is given by $dN/dR \propto R^{-q}$ with $q=3.5$ \citep{Dohnanyi:1969}. A second size distribution would be that measured for solar system comets, normally given as a cumulative size distribution $N(>R) \propto R^{1-q}$. Several investigators have found $q\simeq 2.9$  \citep{Meech:2004, Snodgrass:2011, Fernandez:2013}. \textcolor{black}{A more recent analysis of NEOWISE comet observations by \cite{Bauer:2017} found a debiased value of $q=3.3$ for JFCs similar to the earlier studies, but a shallower debiased $q=2.0$ for LPCs. Note that in comparison with the mass distribution $N(m)\propto m^{-p}$ used in \S\ref{sec:ejection:population}, the power-law exponents are related by $q=(3p-2)$.
}

\textcolor{black}{If we transform these size distributions to differential absolute magnitude distributions $N(H) \propto 10^{\alpha H}$, the exponents of the two forms are related by $\alpha=(q-1)/5$. The above size distributions imply a theoretical magnitude distribution with $\alpha=0.5$, and measured values of $\alpha=0.2\rightarrow 0.46$.} It is clear from Fig.~\ref{fig:iso-sizes} that 1I/`Oumuamua and 2I/Borisov do not match these distributions. 

\textcolor{black}{There are a number of possible explanations. First, they could be from different populations of ISOs, each of which has a more standard size distribution but with significantly different space densities. If they instead reflect the true size distribution of ISOs, this implies the size distribution is steeper than expected. As explained in \S\ref{sec:ejection:population},  ejection processes can produce an increase in the relative number of smaller bodies, consistent with Fig.~\ref{fig:iso-sizes}. An ejection-produced steepening of the size distribution would also explain the mismatch with the observed LPC population that originates from our Oort cloud, where objects have  undergone similar long-term exposure to the ISM but not ejection. Finally, there could be a strong variation in albedo between bodies, possibly due to their individual histories in the ISM or, with 1I/`Oumuamua, by its close perihelion passage pre-discovery. Identifying which of the these explanations is most likely will require further ISO detections for a much better measurement of the magnitude/size distribution.}


\subsection{CNEOS 2014-01-08}
\label{sec:smallISOs}

While 1I/`Oumuamua and 2I/Borisov are {\em macroscopic} ISOs, the existence of interstellar dust particles -- {\em microscopic} ISOs -- has been known for some time \textcolor{black}{from spacecraft detection (see the review by \citealt{Sterken:2019}).} The Advanced Meteor Orbit Radar multi-station complex has provided extensive data on interplanetary dust characteristics. The dynamical information that is collected \textcolor{black}{allows identification of discrete source regions}. One of the main sources appears to be the $\beta$ Pic disk \citep{Baggaley:2000}. During its cruise phase, the {\it Stardust} mission captured particles from the oncoming interstellar dust stream \citep{Westphal:2014}.

In between dust and ISO sizes should lie objects $\sim$1~cm --10~m in size. Fireballs produced by objects  at these sizes  are regularly detected via US DoD satellites, which measure both the luminosity of the fireball and (above a luminosity limit) the entry velocity vector. A search through the publicly accessible fireball catalogue led \cite{Siraj:2022} to identify a potential interstellar impactor, CNEOS 2014-01-08. This object entered the Earths' atmosphere over the Western Pacific  with a reported entry velocity of 44.8 km/s. Although formal measurement uncertainties are not published, DoD personnel communicated that ``the velocity estimate reported to NASA is sufficiently accurate to indicate an interstellar trajectory''. When integrated backwards, the reported velocity vector implied an original orbit with $e=2.4\pm0.3$, hyperbolic at the 3-$\sigma$ level. 

While the identification of an interstellar origin appears sound based on the available satellite data, there is a tension with the fireball and meteor data regularly obtained by ground-based optical and radar meteor surveys. \cite{Musci:2012} identified 2 hyperbolic orbits out of 1739 meteors observed with the Canadian Automated Meteor Observatory, but careful inspection ruled them out as measurement error. Similarly, in an analysis of 824 fireballs detected by the European Fireball Network, \cite{Borovicka:2022} identified two objects that had orbital eccentricities $e>1$ at the 3-$\sigma$ level, but the absolute values were near unity and they concluded there was no strong evidence for a hyperbolic nature. In an analysis of nearly 4000 meteors observed by the FRIPON project, \cite{Colas:2020} report an upper limit of interstellar meteors of $0.1$\% but suspect the true value to be much lower. Collating several meteor orbit catalogues derived from photographic, video and radar systems, \cite{Hajdukova:2020} concluded that there was no convincing evidence of any interstellar meteors being detected in 25 years of data covering many thousands of objects.  \textcolor{black}{That said, a careful analysis of $4\times10^5$ meteors detected by the Canadian Meteor Orbit Radar led \cite{Froncisz:2020} to identify 5 possible interstellar meteors. We note that this possible interstellar fraction of $\sim$1 in 80,000 meteors is significantly lower than the $\sim$1 in 900 implied for the current CNEOS fireball database.}

\textcolor{black}{It is clear from these latter studies that measurement uncertainties are extremely important in interstellar meteor identification, due to the short timespan over which the observations are obtained.} Until such a time that a quantitative description of the uncertainties on the DoD satellite measurements is forthcoming (and indeed their actual values), some uncertainty unfortunately still remains concerning an interstellar origin for CNEOS 2014-01-08. It therefore follows that the pursuit of interstellar meteor detections with quantifiable uncertainties is a highly worthwhile endeavor.


\subsection{ISO impact hazard}

\textcolor{black}{As reported in the Planetary Decadal survey \citep{NAP2022}}, as of 2021 about 95\% of the NEOs greater than 1~km in diameter have been found. These objects are capable of causing global effects upon impact. Objects that are larger than 140~m in diameter can cause regional destruction and to date, only about 33\% of these have been discovered. Estimates show that there might be more than 10$^5$ objects $\ge$50 m in diameter which could cause destruction on urban scales, and less than 2\% of these have been found. In 2005 Congress directed NASA to find 90\% of all NEOs larger than 140m, and to date this directive has not been met. Once it is met, the threat from impacts from near earth asteroids will be minimized, but this will not address the risk from objects on long period Oort cloud comet (OCC) trajectories. It is likely that OCCs are responsible for most of the very large impacts on Earth \citep{Jeffers:2001}.

The typical NEO encounter velocity peaks around 15 km/s, with a range up to 40 km/s \citep{Heinze:2021}, however the encounter velocities for OCCs will typically be around 54 km/s ranging from 16 up to 72 km/s \citep{Jeffers:2001}. ISO trajectories will be very similar to those of OCCs, but the encounter velocities can be even higher. At the time of close approach to the Earth on 2017 Oct. 14, the relative velocity of 1I/`Oumuamua with respect to the Earth was 60.3 km/s. It passed within 63 Earth-moon distances and we did not and could not have discovered it until after it had passed. Objects like 1I/`Oumuamua which are on OCC-like orbits but which do not have any detectable activity, have small nuclei, and low albedo (such as the Manx comets, see \citealt{Meech:2016}) are particularly dangerous because they will be harder to detect. \citet{Heinze:2021} also noted that there is a larger bias than expected against finding small fast moving objects because of the streak length on the detector, so inactive ISOs will be particularly hard to find and we are more likely to be missing more of these.

These OCC and ISO trajectories will be distributed across a wide range of inclinations. As discussed in \S\ref{sec:VCRO}, the powerful new Vera C. Rubin Observatory (Rubin) Legacy Survey of Space and Time (LSST) will begin in mid-2025. There is a concerted effort to optimize the survey effort to benefit solar system science, including detection of OCCs, NEOs and potentially hazardous objects \citep{Schwamb:2023}. 1I/`Oumuamua was discovered at 1.22 au moving at a rate of 6.2$^{\circ}$ per day, i.e. typical of a faint nearby NEO that the LSST will be optimized to detect.  

\section{\textbf{THE NEXT DECADE OF EXOCOMET AND ISO STUDIES -- WHAT DON'T WE KNOW?}}
\label{future_observations}

\subsection{Exocomet Systems}
\subsubsection{Probing outer exocomet populations}

\textcolor{black}{Current and under-construction facilities should lead to significant advances in the field of exocomet science. An important area of study will be to understand the dynamical evolution of exocomets in planetary systems, which requires both high angular resolution and sensitivity at dust-emission wavelengths. Near to mid-IR dust imaging will take place with the 6.5m James Webb Space Telescope (JWST) , the 39-m ESO Extremely Large Telescope (ELT) with its MICADO and METIS cameras \citep{Brandl:2010, Davies:2018},} and large, deep ALMA surveys such as the ongoing ARKS (The ALMA survey to Resolve exoKuiper belt Substructures, Marino et al. in prep.). These will deliver unprecedentedly detailed images of dust emission from exocometary belts, revealing planet-belt interaction and allowing us to infer the dynamical fate of exocomets within belts, be it ejection, inward scattering, or survival within the belt. 

Gas observations offer a unique window into exocometary compositions in young planetary systems just after formation. Near- to mid-IR spectroscopy with JWST combined with ground-based near-IR high-resolution spectrographs (like the newly-installed CRIRES+ instrumentation on the ESO Very Large Telescope and later the ELT+HARMONI) will be used to probe ro-vibrational transitions of new molecules undetectable at mm wavelengths, such as CO$_2$, CH$_4$ and of course H$_2$O and its photodissociation product OH. This will constrain exocometary molecular gas compositions within outer belts, with the potential to confirm the gas origin, determine the physical release mechanism, and ultimately the ice composition of young exocomets - a missing evolutionary link between outer protoplanetary disks and solar system cometary/KBO compositions.

In addition, ALMA CO and C~I line imaging will constrain the radial and vertical distribution and hence the evolution of gas within exocometary belts; this will be combined with detail kinematic analysis, thanks to ALMA's maximum spectral resolution, to reveal the environmental conditions (temperature, bulk density) of the gas, and evaluate the dynamical influence of planets on the evolution of exocometary gas.

\subsubsection{Probing the inward transport of exocomets}

The highest priority for the future of exocomets in the inner regions should be to expand the number of detections and detected systems - currently dominated by $\beta$ Pic - and study their compositions and dynamics as they enter the innermost regions of a planetary system. This will enable us to probe their formation location and the mechanism leading to their inward transport - with the potential of linking their origin to detected exoplanets, and imaged populations of exocomets in outer belts.

JWST \textcolor{black}{is} especially sensitive to warm dust emission in the interior regions of nearby systems, either through its direct imaging or aperture masking modes, and this can be used to set limits on inward scattering and transport of icy exocomets from the outer belt towards the terrestrial region \citep{Marino:2018b}. \textcolor{black}{The Roman Space Telescope is a 2.4-m wide-field optical and near-IR facility, currently planned to launch on 2027.} Its coronagraph instrument is also expected to expand the number of systems with exozodiacal light detections in the terrestrial planet region \citep{Douglas:2022}. Imaging of exozodiacal dust may for the first time also be possible around the nearest stars with the ELT \citep{Roberge:2012}.

\textcolor{black}{The identification and confirmation of more $\beta$ Pic-like systems requires high resolution echelle spectrographs with large aperture telescopes to provide high signal-to-noise per wavelength bin. Observations with new high-resolution spectrographs on the ELT such as ANDES \citep{Maiolino:2013} and METIS \citep{Brandl:2010} will exploit the ELT's large collecting area with their high spectral (R$\sim$100,000) resolution and wide  wavelength coverage. This will allow more sensitive searches for absorption in edge-on systems, both from narrow, stable gas at the stellar velocity and variable red- and blue-shifted features from exocometary gas.} Monitoring observations looking for high-velocity exocomet features should aim to cover volatiles as well as metallic atomic species across a broad wavelength range, \textcolor{black}{as one of the highest priorities is} to understand the nature and volatile content of star-grazing exocomets.

\textcolor{black}{Finally, continued exploration of TESS and CHEOPS stellar light-curves to look for asymmetric exocomet transits are likely to lead to new discoveries. These will be further supported by the PLATO transiting exoplanet mission \citep{Rauer:2014}, with its factor $\sim 5$ increased photometric accuracy over TESS, currently scheduled for launch in 2026. This will contribute to understanding the occurrence rate of star-grazing exocomets around large samples of stars, pushing towards smaller transits.}

\subsection{Galactic evolution of ISOs}

\subsubsection{Galactic effects on ISOs}
\label{sec:ISO_galaxy}

In \S\ref{sec:ejection:population} we saw that ISOs experience various influences from the environment in the form of thermal heating and exposure to radiation and particles of various kinds. The question is whether ISOs also affect the environments they are passing through. The discovery of 1I/`Oumuamua taught us that the ISM contains not only gas and dust but that ISOs are a natural third component of the ISM. 

ISOs also pass through molecular clouds (see section 4.3), with young ISOs more likely to make such an experience than older ones \citep{Pfalzner:2020}. Such a cloud passage might alter not only the ISO surface and structure but also the direction of their path, while at the same time decelerating or accelerating them. Depending on the size of the actual molecular cloud $10^{19}-10^{20}$ ISOs reside in them at any time. Being so small in mass, their individual effect is negligible, but it could be different if co-moving streams of ISOs pass through molecular clouds.

However, although every ISO travels through the ISM, certainly not every ISO passes through a  molecular cloud and even a protoplanetary disk. However, the probability of ISOs passing through molecular clouds is surprisingly high. In the solar neighborhood, an ISO spends $\approx$ 0.1\%-0.2\% of its journey passing through molecular clouds \citep{Pfalzner:2020}. This value increases for young ISOs close to the Galactic center. In comparison, passing through an existing protoplanetary disk is a much rarer event due to its much smaller cross-section. Nevertheless, \citet{Grishin:2019} estimate that at least 10$^4$ ISOs larger than 1 km cross any protoplanetary disk around field stars, with the number increasing to 10$^5$ for star cluster environments.

Under certain conditions, molecular clouds collapse and form stars. It seems that ISOs take part in this process to some extent. This would mean that ISO could become concentrated in collapsing molecular clouds \citep{Pfalzner:2021b}. However, these simulations indicate that there is a competitive capture process at work that favors the capture of ISOs by massive star clusters.

In the current planet formation scenario, there exist some difficulties in proceeding from m-sized to km-sized objects (see \S\ref{sec:planetform:dust_growth}). \textcolor{black}{It has been suggested that ISOs might easily overcome the meter-sized growth barrier by acting as seeds to catalyze planet formation. Two different scenarios have been suggested. \citet{Grishin:2019} theorize that ISOs could be captured from the local star formation region into the disks surrounding young stars . Alternatively, \cite{Pfalzner:2019} propose the ISOs in the ISM taking part in molecular cloud collapse would become a natural component of the forming disk, without the need for additional capture.} They give a first conservative estimate of the order 10$^{11}$ ISOs typically being incorporated into forming star-disk systems. \cite{Moro_Martin:2022} found  a similar total number when considering metre-scale ISOs and larger,  but also noted the number in the disk could increase by 2-3 orders of magnitude in cluster environments.  If ISOs take part in cloud collapse leading to star formation, this would also mean that ISOs are incorporated into forming stars. The question is how many ISOs end up in stars? 

When ISOs pass through molecular clouds and are captured in planet-forming disks, they can be affected by these environments. Possible mechanisms are devolatilization and erosion. Both will lead to changes in the ISO size distribution. Whether the ISOs are altered depends primarily on the gas/dust density, which increases from an ISM environment to molecular clouds and protoplanetary disks. 

Finally, \cite{Do:2018} point out that exo-Oort clouds around stars that produce supernovae will be irradiated and lose surface volatiles, but the exocomets there will survive. As they then drift away from the supernova remnant they will form a natural population of devolatilized ISOs, possibly somewhat similar to 1I/`Oumuamua. However supernovae are rare events, and such ISOs should not form a major component of the population.

\subsubsection{
ISO effects on the galaxy}
\label{sec:ISO_transport}

Over the last four years, it has become apparent that ISOs are a natural component of molecular clouds. The question is whether the presence of the huge number of low-mass ISOs does ``contaminate'' star forming regions. This depends on the mass of the ISOs present in molecular clouds, but also on the different chemistry between  star forming  systems and on whether ISOs are evenly spread or come in streams from their parent systems. 

So far this question has been studied more from the perspective of individual already forming planetary systems. This question of whether material can be spread from one planetary system to another is of long standing interest \citep{Valtonen:1982,Melosh:2003,Adams:2005, Brasser:2006}. More recently, estimates of the likelihood of ISO capture events have been performed for the early phase in young star-forming clusters and the local galactic neighborhood \citep{Hands:2019,Portegies:2021,Napier:2021}. In young clusters the stellar density is much higher than in the field, therefore, the ISO capture rate per Myr is naturally much higher. However, whether the capture probability is in general higher during the short cluster phase than over the entire lifetime of a star depends strongly on the assumed cluster properties and the stellar mass. The latter determines the capture cross section and the lifetime of a star. Besides, GAIA data strongly indicate that clusters seem not completely dissolved within 10 Myr, but move together in an unbound state for a considerable length of time. Thus assuming field densities for the stars and the ISOs underestimate the capture rates in the first few hundred Myr. This requires further investigation in the future.

On average every star will contribute ISOs to the galactic population, and this material might be injected into the inner regions of star systems. \citet{Seligman:2022b} suggest that this material can impact both the host stars and their planets, enriching their atmospheres, and that understanding the composition of the ISO population will have implications for post-formation exoplanet atmospheric composition.

Currently there exists no definitive evidence that any gravitationally bound objects in the solar system are of extrinsic origin \citep{Morbi:2020}.  However, the discovery of 1I/`Oumuamua and 2I/Borisov sparked wide-ranging speculation regarding the possibility that our solar system is being more broadly contaminated by minor bodies of extra-solar origin \citep{Siraj:2019,Namoui:2020,Hands:2020}.
There exists a wide range in the estimated masses of planetesimals accreted from other stars while the Sun lived within its birth cluster from 10$^{-5}\mearth$  to a third of the Oort cloud population. \cite{Napier:2021} finds that about $10^6$ more ISOs were captured during the cluster phase than accreted subsequently from the field, while the current steady state factor is $10^4$ and the total mass of surviving captured ISOs is $\sim 10^{-9}\mearth$. To put this into perspective, the mass of the Oort cloud is estimated to range from 0.75 $\pm$ 0.25 $\mearth$ \citep{Brasser:2008} to 1.9 $\mearth$ \citep{Weissman:1996}. With a typical comet-mass of a few times 10$^{12}$ to 10$^{14}$ kg \citep{Sosa:2011} the Oort cloud contains approximately 10$^{10}$-10$^{12}$ comet-sized objects. Thus the captured ISO population is probably extremely small in comparison to the primordial comet population in the Oort cloud. 

Current capture of ISOs into the solar system has also been explored. \citet{Dehnen:2022} suggest that planets capture $\sim$2 ISOs every 1000 years which would result in 8 ISOs captured within 5 au of the Sun at any time.   A perennial candidate for a captured ISO is comet 96P/Machholz, due to its highly anomalous optical spectrum that is dominated by NH$_2$, with strong depletions of carbon-based molecules \citep{Langland-Shula:2007, Schleicher:2008}. Its orbital elements of $q=0.124$, $i=58^\circ$ and $e=0.959$ mark it as one of the lowest perihelia comets known, although its high inclination is due to strong Kozai-Lidov oscillations \citep{McIntosh:1990}. Its nucleus was studied by \cite{Eisner:2019} to compare with 1I/`Oumuamua to search for possible effects of small perihelion passages. The strongest evidence for it being a captured ISO remains its peculiar coma composition, but this is not a definitive marker.

\subsubsection{Intergalactic Objects}
\label{sec:Intergalactic}

The existence of ISOs in the Milky Way makes it likely that other galaxies are also populated with large numbers of ISOs. One can also speculate about the number of  objects that have been ejected from their parent galaxy and become Intergalactic Objects.  These would clearly be difficult to create, given the huge velocities required; at the Sun's position  the escape velocity from the Milky Way is $\simeq 500$ km/s \citep{Koppelman:2021}. While some tens of hypervelocity stars escaping the Milky Way have now been identified \citep{Li:2021}, most are thought to have attained their velocities via encounters with massive black holes, whose dynamical and radiation environments are hardly conducive for ISO survival. 

However, we note that while escape to intergalactic space is highly unlikely for an ISO, we can let the galaxy carry it to a new home. Galactic mergers have taken place throughout the history of the Universe, with 5 minor mergers identified for the Milky Way \citep{Kruijssen:2020}. An unknown fraction of the ISOs in those galaxies would have merged with the galactic population, and of course subsumed stellar systems would have ejected their own ISOs through the usual methods described in \S\ref{sec:ejection}. Hence given enough time as ISO detections increase, there is a finite but as-yet unquantified probability of detecting an ISO carried here by its own galaxy.

\subsection{The range of ISO properties}
\subsubsection{Future detections and surveys}
\label{sec:VCRO}

\begin{figure}[ht!]
\begin{center}
\includegraphics[width=8.3cm]{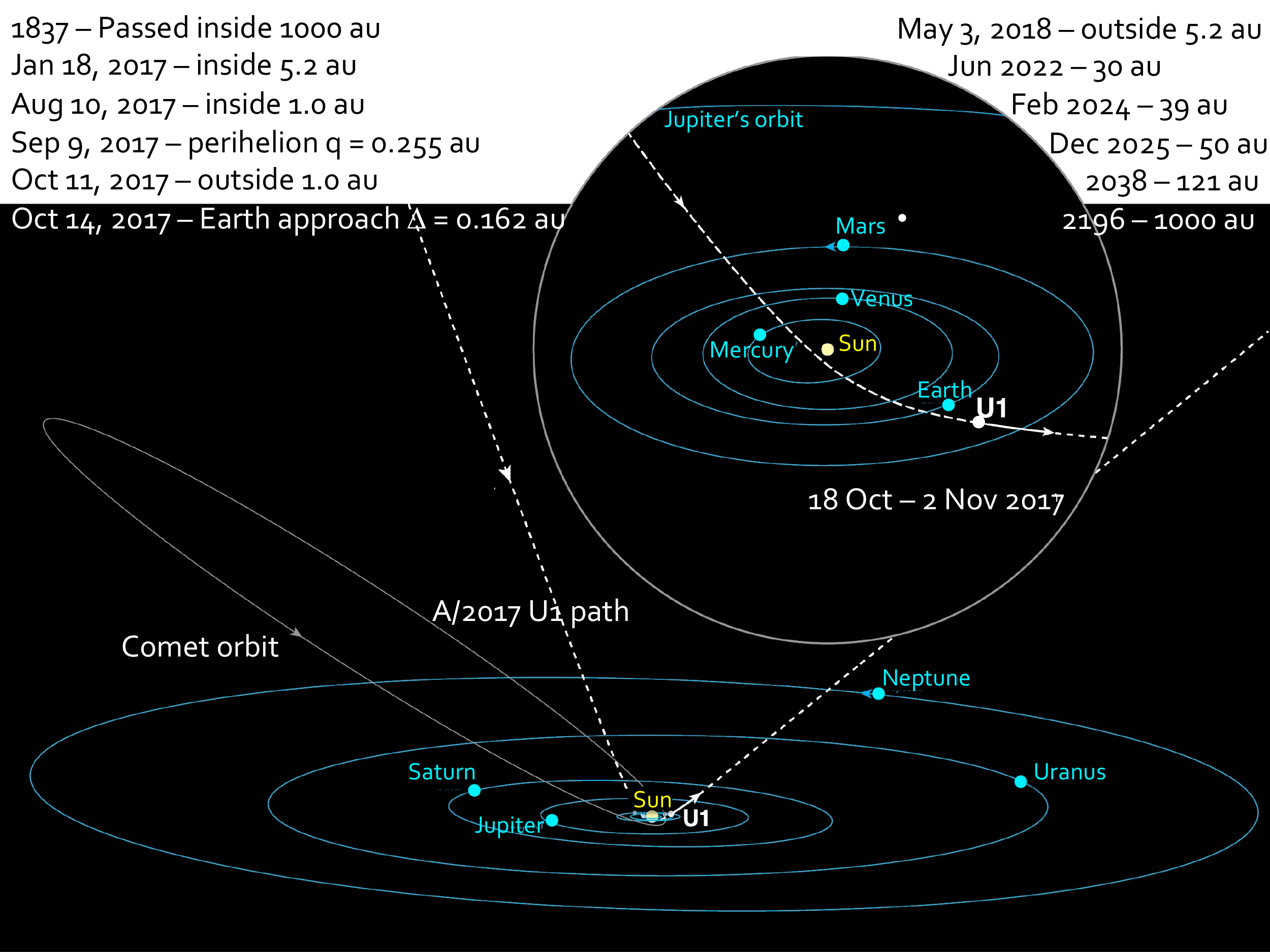}
\caption{Path of 1I/`Oumuamua as it entered the solar system illustrating why it was not discovered sooner.}
\label{fig:path}
\end{center}
\vspace{-0.25cm}
\end{figure}

The trajectory of 1I/`Oumuamua brought it in from above the plane of the solar system from the direction of the constellation Lyra (Fig.~\ref{fig:path}). It came to perihelion
inside the orbit of Mercury at $q$ = 0.255 au on 2017 Sep. 9. and was discovered post-perihelion at $r$ = 1.22 au after it had made its close approach to Earth on 2017 Oct. 14. This close approach was at 0.162 au, passing within 63 lunar distances. Images from the Catalina Sky Survey detected it on Oct. 14, but it was noted only after it had been discovered. 

\textcolor{black}{\Ou~passed through the SOHO and STEREO fields of view in early 2017 Sep., near perihelion \citep{Hui:2019}. Because of the large phase angle and the extreme forward scattering, this would have enhanced the brightness of any dust around the nucleus by $\sim$ two orders of magnitude but neither satellite detected it}. It could not have been discovered any earlier than it was because 1I/`Oumuamua was just coming out of solar conjunction as seen from the ground and was faint (mag$\sim$23.7 at 0.79 au on Oct. 2). Prior to moving into solar conjunction in early August it would have been fainter than mag$\sim$25 - both times too faint to be discovered by any of the existing surveys. 1I/`Oumuamua was inside the orbit of Jupiter for less than 1.3 yrs. 
The second ISO was discovered by an amateur searching close to the horizon in twilight -- something that the large telescope surveys cannot do (see Bauer {\it et al.}, this volume). 

The Rubin observatory in Chile has an 8.3m survey telescope with a camera that covers a FOV of 9.6 deg$^2$. It will begin regular survey operations in mid-2025 that will provide a much greater depth than all previous surveys, reaching mag$\sim$24.7 and scanning the accessible sky every 3 days \citep{Jones:2009}. \citet{Hoover:2022} have estimated the number of discoveries that might be made by the LSST by generating a synthetic population of ISOs, assuming no cometary activity.  They predict that the survey will find on the order of 1 ISO per year, but the number can be larger than 1 when cometary activity is considered. \textcolor{black}{Ironically, however, even had the LSST survey been operational in 2017, it would have been unlikely to discover 1I/`Oumuamua because the ISO would have had an {\it average} brightness brighter than mag$\sim$24.7 for less than a week before moving into solar conjunction in early 2017 Aug.}

\subsubsection{Probing origins and evolution}
\label{sec:origins}

The standard mechanisms of ISO creation via ejection of exocomets from their home systems are described in detail in \S\ref{sec:ejection}. 1I/`Oumuamua's discovery resulted in the proposal of additional possible creation routes. Many of the formation mechanisms have been suggested in an attempt to explain the origin of the unusual shape for 1I/`Oumuamua. These include fluidization to a Jacobi ellipsoid during the red giant phase of a star \citep{Katz:2018}, interstellar ablation \citep{Vavilov:2019}, collisional elongation \citep{Sugiura:2019}, formation as an icy fractal aggregate \citep{Moro-Martin:2019b} and a tidally disrupted planetesimal that passed close to growing giant planets, from which the volatiles were removed during a close stellar passage \citep{Raymond:2018b, {Zhang:2020}}.

\textcolor{black}{Two \Ou~origin scenarios suggested a volatile composition based on homonuclear diatomic molecules with no dipole moments and no vibrational spectral lines, which would explain why no gas was detected}. In an attempt to both explain the shape and acceleration of 1I/`Oumuamua, \citet{Seligman:2020} suggested that it was composed of molecular hydrogen ice. In their model, mass wasting from sublimation far out in the solar system could explain both the shape and the acceleration. They proposed that the ISO formed in a cold dense molecular cloud core. The freezing point of H$_2$ is around 14K, and the lowest temperature cloud cores are around 10K. \textcolor{black}{H$_2$ has no dipole moment and can only be detected in the far-UV or in the infrared through its rotational lines from space.} An alternate suggestion was made by \citet{Desch:2021} who proposed that 1I/`Oumuamua was a collisional fragment from an exo-Kuiper belt.  Many large KBOs in our solar system exhibit N$_2$ ice on their surfaces.  Gaseous N$_2$ has no bending mode vibration spectral lines and is infra-red inactive.  Its spectral features are in the UV where no observations were taken. This makes it an attractive volatile to explain how 1I/`Oumuamua could have unobserved outgassing. \citet{Levine:2021} suggests that neither scenario is plausible because large H$_2$-ice bodies are not likely to form in cloud cores and that the impacts in a Kuiper belt are not sufficient to generate large N$_2$ fragments.

\begin{figure*}[ht!]
\begin{center}
\includegraphics[width=14.5cm]{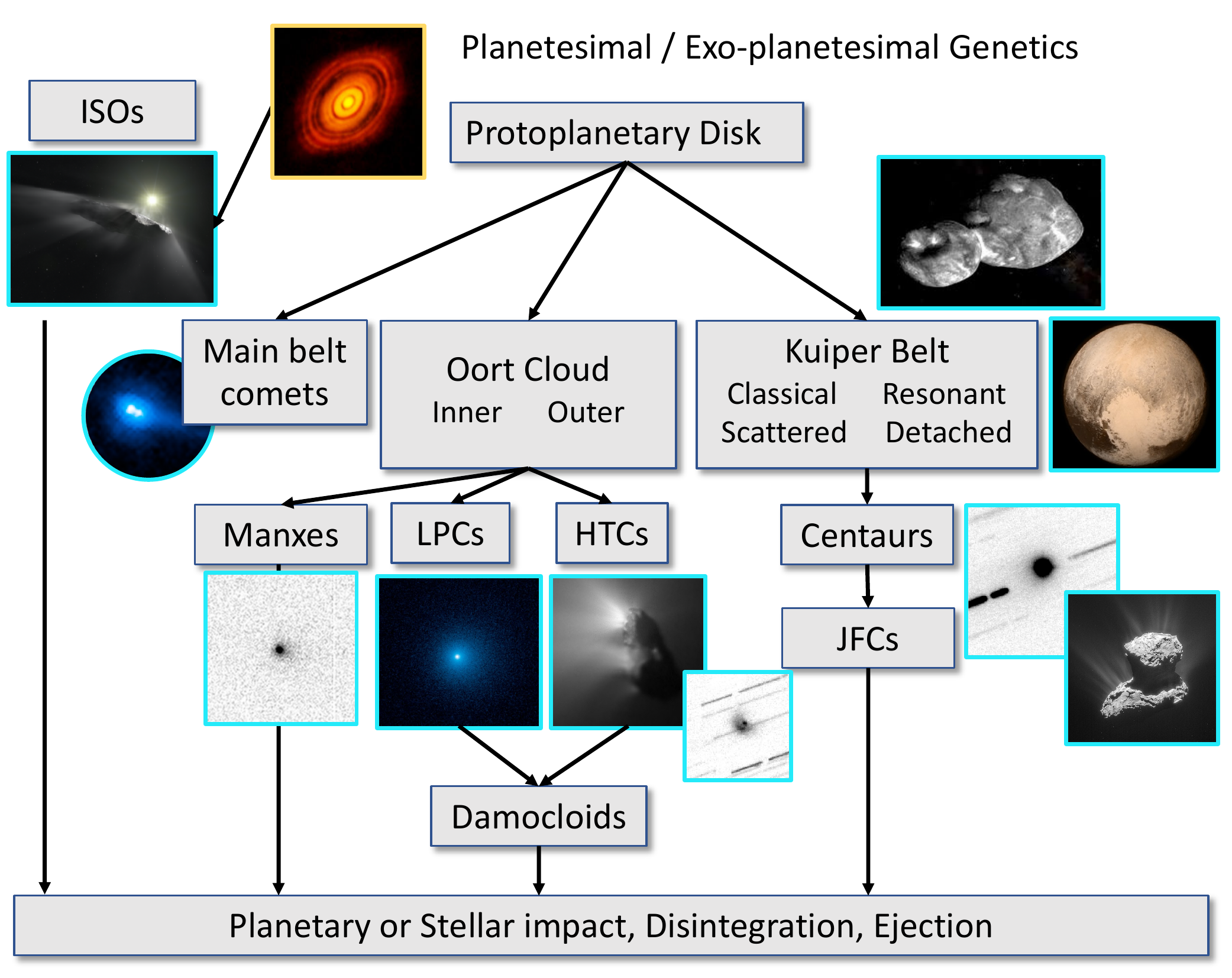}
\caption{Genetic relationships of early solar system planetesimals both in our solar system and from exoplanetary systems. We are just at the beginning of exploring the ISO population. Objects on OCC orbits, including ISOs, represent the largest reservoir of objects that has not been explored with an in-situ mission. HTC = Halley Type comet, JFC = Jupiter Family Comet).}
\label{fig:Summary}
\end{center}
\end{figure*}

\textcolor{black}{In order to reconcile some of the observational and model inconsistencies, \citet{Bergner:2023} propose that the acceleration of 1I/`Oumuamua is caused by release of molecular hydrogen that formed by cosmic ray processing of water ice. These authors argue that laboratory experiments show that H$_2$ is efficiently produced in water ice during radiation by high energy particles. They propose that the H$_2$ gas is released over a wide range of temperatures during annealing of amorphous water ice and that there is sufficient gas released to account for the observed non-gravitational acceleration.}

Finally, regarding origin location within a disk, \citet{Seligman:2022} propose that measuring the C/O ratio of an ISO can be a tracer of whether the ISO formed inside or outside the ice line in its home star system.

\subsection{In-situ Observations: Space missions}
\label{sec:in-situ}

1I`/Oumuamua was accessible to ground based telescopes for less than a month, a little longer using space facilities. After this brief period of observation it was found that the characteristics were quite different from what was expected from the first ISO, and this one discovery has energized a new interdisciplinary awareness in the study of planet formation. However, a more detailed investigation of ISOs with an in-situ mission presents unique challenges: the orbits may have high inclinations and the encounter speeds are typically high (10s of km/s) which means a short encounter. \textcolor{black}{There are increased risks with high velocity flybys if the ISO ejects large dust particles. The Giotto spacecraft was destabilized with a cm-sized impactor at a relative velocity of 68 km/s \citep{Bird:1988}}.

NASA’s competed mission calls are not compatible with missions that are responsive to new discoveries, i.e., missions that do not have a target at the time they are proposed. This is relevant for ISOs as well as OCCs, Manx comets, and hazardous NEOs. Two approaches have been suggested to explore these targets: spacecraft in storage, ready to launch following target discovery and spacecraft in a standby orbit, as is being done by ESA’s Comet Interceptor mission \citep[see Snodgrass {\it et al} this volume, and][]{Snodgrass:2019, Sanchez:2021}. Typically, the target will be unknown at the time of spacecraft development and this has an effect on the definition of basic spacecraft capability (e.g., $\Delta$v) and on the payload. Launch following the discovery of an ISO offers a greater flexibility in terms of target access but requires a very fast turnaround of a launch vehicle. A spacecraft in a standby orbit is more responsive to a target but has a more limited target accessibility.  

Using known OCC orbits, an assessment of the phase space of targets with available launch energies, maximum encounter speed and for times of flight of less than 10 years, launching while the comet is more than 0.5 years from coming to perihelion, shows that for low launch energies, C$_3$, there are almost no targets. {\textcolor{black}{(In astrodynamics,  the characteristic energy C$_3$ km$^2$/s$^2$ is the measure of the excess specific energy over that required to just barely escape a massive body; C$_3$ = v$_{\infty}^2$}.)} There are also almost no targets unless the encounter speeds are greater than 10 km/s.  This suggests that at present only fast flyby missions are an option for ISOs.  For a comprehensive mission, this likely requires significant advancements in payload capabilities (e.g. small deployable satellites and autonomous navigation) \citep{Donitz:2021, Donitz:2022}.

\textcolor{black}{There were several concepts developed for missions that could reach \Ou. \citet{Seligman:2018} estimated that for launch-on-detection scenarios, the wait time would be on the order of 10 years for a favorable mission opportunity. \citet{Hibberd:2020} explored a range of flyby trajectories which could launch in the early 2030’s and deliver a spacecraft to \Ou~with relative velocities between 15-20 km/s, arriving between 2048-2052. Finally, \citet{Miller:2022} proposed high-performance solar sail scenarios that would allow for a rendezvous with an ISO by maintaining a high potential energy position until the detection of an ISO and then matching velocity through a controlled fall toward the Sun. }

\textcolor{black}{The LSST survey will increase the number of discoveries by going much fainter, and therefore to larger discovery distances. As shown by \citet{Engelhardt:2017}, ISOs are more common at larger perihelion distances. The fainter limiting survey magnitude would enable detection of a 1 km inactive nucleus with 4\% albedo out to 5.5 au, active ISOs much further.  Although most detected ISOs may be too distant to easily reach with spacecraft, \Ou~and 2I/Borisov demonstrate ISOs exist that come within 2 au, and detection of an incoming ISO at large distances may provide enough warning time for either the storage or standby concepts discussed above.
}

\section{\textbf{SUMMARY }}
\label{sec:summary}

From the multitude of studies described in this chapter, it is clear that exocomets should be and are common in nearby stellar systems. Together with the physical mechanisms by which they can be lost to interstellar space, the local stellar neighborhood should be rich in ISOs, a prediction finally confirmed through the discovery of 1I/`Oumuamua and 2I/Borisov.
So the question remains, how is 1I/`Oumuamua not visibly active while 2I/Borisov is strongly active; is this due to origin, age, or the unseen 1I/`Oumuamua perihelion passage? From models of our solar system \citep{Shannon:2015}, some fraction of ISOs will not be cometary but more like C- or S-type asteroids. Based on the observation of Manx comets \citep{Meech:2016}, we should be able to distinguish asteroidal objects from inert comets. Is there a continuum of properties for ISOs, or have we already found two separate populations arriving from different origin mechanisms? This sample of two very different ISOs makes it difficult to predict what will 3I be like - will it be like planetesimals from our solar system or something different (see Fig.~\ref{fig:Summary})? Given the increasing sensitivity of sky surveys, the current and next generation of optical, IR and sub-mm facilities, plus the theoretical advances in cometary/exocomet structure and evolution, these open questions have a chance of being answered in {\it Comets IV}.\\

\noindent \textbf{Acknowledgments} \smallskip

We thank the reviewers, Matthew Knight, Sean Raymond and an anonymous referee for the very thorough and helpful reviews on a very short timescale! K.J.M. acknowledges support through NASA Grant 80-NSSC18K0853 and in addition to support for HST programs GO/DD-15405, -15447, 16043, -16088, and -16915 provided by NASA through a grant from the Space Telescope Science Institute, which is operated by the Association of Universities for Research in Astronomy under NASA contract NAS 5-26555. A.F. acknowledges support from UK STFC award ST/X000923/1. Figure \ref{fig:betapic:comets} is based on data obtained from the ESO Science Archive Facility with DOI  https://doi.org/10.18727/archive/33 .

\bibliographystyle{sss-three.bst}
\bibliography{refs.bib}

\end{document}

%% file: tab-1I-2I.tex
\begin{table*}
\begin{center}
\caption{A summary of measured properties of interstellar objects \label{tab:1I-2I}}
\begin{tabular}{lllclc}
\hline
Quantity                   &                                     & 1I/`Oumuamua                             & Ref$^{\ddag}$   & 2I/Borisov                   &  Ref$^{\ddag}$ \\
\hline
\multicolumn{3}{l}{\bf Dynamical Properties} \\
Perihelion date            & $T_p$                               &   2017 September 09.51                   & [1]   & 2019 December 08.55 
         & [11] \\
Perihelion distance        & $q$ [au]                            &   0.256                                  & [1]   &  2.007                                 & [11] \\
Eccentricity               & $e$                                 &   1.201                                  & [1]   &  3.356                                 & [11] \\
Earth close approach       & $\Delta$ [au]                       &   0.162                                  & [1]   &  1.937                                 & [11] \\
Incoming velocity$^{\dag}$ & $v_{\infty}$ [km s$^{-1}$]          & $26.420\pm0.002$                         & [2]   & 32.288                                 & [11] \\
Non-grav acceleration      & $A_1$ $\times$ 10$^8$ [au d$^{-2}$] & 27.90                                    & [1]   &  7.09                                  & [11] \\       
Non-grav acceleration      & $A_2$ $\times$ 10$^8$ [au d$^{-2}$] &  1.44                                    & [1]   & -1.44                                  & [11] \\       
Non-grav acceleration      & $A_3$ $\times$ 10$^8$ [au d$^{-2}$] &  1.57                                    & [1]   &  0.065                                 & [11] \\       
\hline
\multicolumn{3}{l}{\bf Physical Properties} \\
Absolute magnitude         & $H_V$                               & $22.4\pm0.04$                            & [3]   & $>$19.1                                  & [12] \\
Albedo                     & $p_V$                               & 0.01 -- 0.2                              & [4]   & --                                     &      \\
Radius (for $p_V$=0.04)    & $r_N$ [m]                           & 110                                      & [3]   & $<$ 500                                & [12] \\
Rotation state             &                                     & complex, long-axis mode                  & [5]   & --                                     &      \\ 
Rotation period            & $P$ [hr]                            & $8.67\pm0.34$~hr                         & [5]   & --                                     &      \\
Axis ratio                 & a:b                                 & $>$6:1                                   & [6]   & --                                     &      \\
Spectral slope             & $S_V$ [\% per 100 nm]               & (7-23)$\pm$3                             & [6,7] & 12$\pm$1                               & [8]  \\ 
H$_2$O production          & $Q$(H$_2$O) [molec s$^{-1}$]        &\textcolor{black}{$<6.1   \times$10$^{25}$ @ 0.38 au (obs)}  & [9]   & 1.1 $\times$10$^{27}$ @ 2.0 au  (obs)  & [13] \\
H$_2$O production          & $Q$(H$_2$O) [molec s$^{-1}$]        & 4.9   $\times$10$^{25}$ @ 1.4 au (model) & [10]  & --                                     &      \\
Hyper volatile (CO?)       & $Q$(X) [molec s$^{-1}$]             & 4.5   $\times$10$^{25}$ @ 1.4 au (model) & [10]  & --                                     &      \\
CO$_{2}$ production        & $Q$(CO$_2$) [molec s$^{-1}$]        & $<$ 9 $\times$10$^{22}$ @ 2.0 au (obs)   & [4]   & --                                     &      \\
CO production              & $Q$(CO) [molec s$^{-1}$]            & $<$ 1 $\times$10$^{24}$ @ 2.0 au (obs)   & [4]   & 4.4 $\times$10$^{26}$ @ 2.0 au (obs)   & [14] \\
Dust production            & $Q$(dust) [kg s$^{-1}$]             & $<$ 10$^{-3}$ @ 1.4 au (obs)              & [3,7] &  $2-50$ @ 2.6 au (obs)                & [15,16]\\
\hline
\end{tabular}
\end{center}
$^{\dag}$ \textcolor{black}{$v_{\infty}$ is the velocity an object on a hyperbolic orbit has infinitely far from the central body (here, the Sun)}; $^{\ddag}$Reference Key: [1] Osculating orbit: JPL Horizons orbital solution \#16 [2] speed relative to the Sun while in interstellar space \citep{Bailer-Jones:2018}; [3] \citet{Meech:2017}; [4] the CO production rate is corrected from the number reported by \citet{Trilling:2018}; [5] \citet{Belton:2018}; [6] \citet{ISSIteam:2019}; [7] \citet{Jewitt:2017}; [8] \citet{Jewitt:2022};  [9] \textcolor{black}{\citet{Hui:2019};} [10] \citet{Micheli:2018}; [11] JPL Horizons orbital solution \#53, [12] \cite{Jewitt:2020a}, [13] \cite{Xing:2020}, [14] \cite{Cordiner:2020}, [15] \cite{JewittLuu:2019}, [16] \cite{deLeon:2019}.  For production rates of other species for 2I/Borisov, see the summary in \citet{Jewitt:2022}.  
\end{table*}